\documentclass{article}
\usepackage{float}
\usepackage{amssymb}
\usepackage{amsmath}
\usepackage{amsfonts}
\usepackage{graphics,graphicx,epsfig}
\usepackage{color}
\usepackage{bm}      
\usepackage{amsfonts} 
\usepackage{caption}
\usepackage{subcaption}
\usepackage{adjustbox}
\usepackage{xcolor}
\usepackage[font=small]{caption}
\usepackage[english]{babel}

\begin{document}

\title{Generalized confidence intervals for the covariate-specific overlap coefficient}
\author{A. Alcaraz-Peñalba$^{1,2}$, A.M. Franco-Pereira$^{2,3},$ M.C. Pardo$^{2,3}$\\
\small%
$^{1}$University Hospital 12 de Octubre, 28041-Madrid, Spain\\
\small%
$^{2}$Department of Statistics and O.R., Complutense University of Madrid, 28040-Madrid, Spain\\
\small%
$^{3}$Instituto de Matem\'{a}tica Interdisciplinar, Complutense University of Madrid, 28040-Madrid, Spain
}
\date{}
\maketitle

\begin{abstract}
Continuous diagnostic tests are often used to discriminate between diseased and healthy populations. Therefore, it is important to look for measures to assess the effectiveness of diagnostic markers in discriminating between these two states. The overlap coefficient (OVL), which is defined as the area of intersection of the graphs of two probability density functions, is considered as a summary measure of diagnostic accuracy. Often additional information is available on some of the covariates which are known to influence the accuracy of such measures. Consequently, in this work we incorporate covariate information using linear regression models with normally distributed errors, enabling the construction of the OVL estimator based on the ordinary least squares estimates of the regression coefficients. We further develop a method for constructing generalized confidence intervals for the covariate-specific overlap coefficient. A simulation study is conducted to assess the finite-sample performance of the proposed generalized confidence interval method and to compare it with other procedures. Finally, an illustration to assess the age-specific accuracy of glucose as a biomarker of diabetes is provided.
\end{abstract}

\noindent \underline{\textbf{Keywords and phrases}}\textbf{:} Overlap coefficient (OVL); covariate adjustment; generalized pivotal quantity; linear regression models.

\section{Introduction}
The overlap measure, also known as overlap coefficient, is already a well-studied quantity originally introduced by Weitzman (1970). It can be used to assess similarity of distributions in several fields as referenced in Franco-Pereira et al. (2021) and Pardo et al. (2025). The overlap coefficient defined for two populations, corresponds to the common area shared by two density functions. Mathematically, given two continuous random variables $X_1$ and $X_2$ representing measurements in two groups, with densities $f_{X_1}(x)$ and $f_{X_2}(x)$, the OVL is defined as:
\begin{equation*}
\text{OVL} = \int \min\{f_{X_1}(x), f_{X_2}(x)\}\,dx
\label{def OVL 1}
\end{equation*}
This integral captures the region of overlap between the two distributions, regardless of whether they follow the same parametric form. The OVL ranges from 0 to 1, where smaller the overlapping area, the better the biomarker is at discrimination.
\newline

Samawi et al. (2017) provides a graphical demonstration and mathematical derivation of the relationship between the Youden index J and the overlap coefficient OVL. They show that J is equal to 1-OVL if the densities of the healthy and diseased populations have only one intersection but OVL and J are not equivalent in other cases. That is to say, under normality this happens if the variances are equal. Furthermore, the Youden index only captures the maximum distance of the CDFs at the single optimal threshold. However, OVL captures all differences at each of the intersections. Pardo and Franco (2026) explored several overlap-based measures as alternatives to classical ROC summary statistics. Their study revealed connections between overlap measures and statistical distances and identified scenarios in which these measures outperform ROC-based metrics. Therefore, overlap measures can be more informative than summary indices such as \(J\) in scenarios where differences between the two distributions are not adequately captured by a single optimal threshold.
\newline

Smith (1982) derived formulas for estimating the mean and variance of a discrete version of Weitzman’s measure using the delta method. Mishra et al. (1986) provided small- and large-sample properties of the sampling distribution of a function of the OVL under the assumption of homogeneity of variances. Mulekar and Mishra (1994) simulated the sampling distribution of estimators of overlap measures when the two densities correspond to the normal case with equal means, and obtained approximate expressions for the bias and variance of these estimators. Numerous works have investigated methods for inference on the OVL (Al-Saidy et al., 2005; Al-Saleh and Samawi, 2007; Samawi and Al-Saleh, 2008; Chaubey et al., 2008; Helu and Samawi, 2011; Reiser and Faraggi, 1999; Samawi et al., 2011; Mulekar and Mishra, 2000; Mizuno et al., 2005; Clemons and Bradley, 2000; Schmid and Schmidt, 2006; Anderson et al., 2012). More recently, Wang and Tian (2017), Franco-Pereira et al. (2021), and Alcaraz-Peñalba et al. (2026) proposed several methods for constructing confidence intervals for the OVL.
\newline

The effectiveness of a continuous marker can be influenced by covariates/factors such as age, gender, general health status, etc. When covariates are not considered, the diagnostic accuracy of the test can be biased or misleading. By incorporating information from covariates using linear regression models, Hu et al. (2021) generalized confidence intervals for the covariate-adjusted Youden index and its optimal cut-off point. For OVL, covariate effects have been adjusted for constructing confidence intervals (Pardo et al., 2025). However, generalized confidence intervals have never been proposed for this overlap measure. As pointed before only Youden index and overlap measure are related when only the densities have one intersection so it is worthwhile to study this method of intervals construction for OVL. The concepts of generalized confidence interval have been successfully applied to a variety of practical settings where standard exact solutions do not exist for confidence intervals and hypothesis testing. It has been shown that generalized inference approaches typically have good performance, even at small sample sizes; for example, see Weerahandi and Berger (1999), Krishnamoorthy and Lu (2003), Lin et al. (2007), Wang and Tian (2017) and Alcaraz-Peñalba et al. (2026).
\newline


The rest of the paper is organized as follows. In Section 2, we model the test results of a biomarker as a function of covariates, employing linear regression model for healthy and disease groups separately. An estimator of the covariate-specific OVL is derived from the ordinary least squares estimates of the regression model. In Section 3, an algorithm for computing the generalized confidence intervals for the covariate-specific OVL is outlined. Section 4 reports the results of simulation studies evaluating the coverage probabilities and average widths of the proposed generalized confidence interval and other known methods. In Section 5, our method is illustrated using data from a population based survey where the goal is to infer about the age-specific accuracy of the glucose as a biomarker of diabetes. Concluding remarks are offered in Section 6.

\section{OVL estimation under a linear regression model for the biomarkers}
Let $Z_1,\ldots,Z_p$ be continuous random variables representing the $p$ covariates, and let $\mathbf{Z} =(1,Z_1,\ldots,Z_p)$ be the augmented random vector including an intercept. Let $\mathbf{z} =(1,z_1,\ldots,z_p) \in \mathbb{R}^{p+1}$ denote a fixed value of this vector. We assume that the test results in Group 1 and Group 2, denoted by $X_1$ and $X_2$, respectively, are modeled as linear functions of the covariates. Specifically, for the random variable associated with Group 1:

\begin{equation}
X_1 \mid (\mathbf{Z} = \mathbf{z}) = \mathbf{z} \bm{\beta}_1^T + \varepsilon_1
\label{eq:LM1}
\end{equation}

\noindent Analogously, for the random variable associated with Group 2:

\begin{equation}
X_2 \mid (\mathbf{Z} = \mathbf{z}) = \mathbf{z} \bm{\beta}_2^T + \varepsilon_2
\label{eq:LM2}
\end{equation}

where, for $i = 1, 2$, $\bm{\beta}_i^T$ denotes a $(p+1) \times 1$ column vector of regression coefficients, and
$\varepsilon_i \sim \mathcal{N}(0,\sigma_i^2)$ represents the corresponding error term. The error terms are assumed to be independent of the covariates and of each other. 
\newline

Under the assumptions of models~\eqref{eq:LM1} and~\eqref{eq:LM2}, the random variables \(X_1 \mid (\mathbf{Z} = \mathbf{z})\) and \(X_2 \mid (\mathbf{Z} = \mathbf{z})\) are independent and follow normal distributions \(\mathcal{N}(\mu_1( \mathbf{z}), \sigma_1^2)\) and \( \mathcal{N}(\mu_2(\mathbf{z}), \sigma_2^2)\), respectively, being \(\mu_i(\mathbf{z}) = \mathbf{z} \bm{\beta}_i^T\), \(i = 1, 2\). Considering \(\sigma_1 < \sigma_2\), the overlap coefficient between \(X_1 \mid (\mathbf{Z} = \mathbf{z})\) and \(X_2 \mid (\mathbf{Z} = \mathbf{z})\) can be computed using the parametric formula for the binormal model (Inman, 1984; Mulekar and Pereira De Andrade, 2017):

\begin{equation}
{OVL}(\mathbf{z}) = 1 + \Phi\left(\frac{x_1(\mathbf{z}) - \mu_1(\mathbf{z})}{\sigma_1}\right) 
                        - \Phi\left(\frac{x_1(\mathbf{\mathbf{z}}) - \mu_2(\mathbf{z})}{\sigma_2}\right)
                        - \Phi\left(\frac{x_2(\mathbf{z}) - \mu_1(\mathbf{z})}{\sigma_1}\right) 
                        + \Phi\left(\frac{x_2(\mathbf{z}) - \mu_2(\mathbf{z})}{\sigma_2}\right),
\label{eq:OVLz1}
\end{equation}

where
\begin{equation*}
x_1(\mathbf{z}) = \frac{ \mu_1(\mathbf{z}) \, \sigma_2^2 - \mu_2(\mathbf{z}) \, \sigma_1^2 
                  - \sigma_1 \sigma_2  
                    \sqrt{ \left(\mu_1(\mathbf{z}) - \mu_2(\mathbf{z})\right)^2 
                           + \left( \sigma_1^2 - \sigma_2^2 \right) 
                             \log\left( \frac{ \sigma_1^2 }{ \sigma_2^2 } \right) } }
                  { \sigma_2^2 - \sigma_1^2 },
\label{eq:x1z1}
\end{equation*}

\begin{equation*}
x_2(\mathbf{z}) = \frac{ \mu_1(\mathbf{z}) \, \sigma_2^2 - \mu_2(\mathbf{z}) \, \sigma_1^2 
                  + \sigma_1 \sigma_2  
                    \sqrt{ \left(\mu_1(\mathbf{z}) - \mu_2(\mathbf{z})\right)^2 
                           + \left( \sigma_1^2 - \sigma_2^2 \right) 
                             \log\left( \frac{ \sigma_1^2 }{ \sigma_2^2 } \right) } }
                  { \sigma_2^2 - \sigma_1^2 },
\label{eq:x2z1}
\end{equation*}

and $\Phi$ denotes the cumulative distribution function of the standard normal distribution. When $\sigma_2 < \sigma_1$, it suffices to interchange
the group labels to obtain the covariate-specific overlap coefficient. If $\sigma_1 = \sigma_2 = \sigma$ and $\mu_1(\mathbf{z}) \neq \mu_2(\mathbf{z})$, the two conditional normal densities have a unique intersection point given by
\[x(\mathbf{z}) = \frac{\mu_1(\mathbf{z})+\mu_2(\mathbf{z})}{2}.\]

In the equal-variance case, the overlap coefficientis given by
\begin{equation}
OVL(\mathbf{z})=2\Phi\left(-\frac{|\mu_1(\mathbf{z})-\mu_2(\mathbf{z})|}{2\sigma}\right).
\label{eq:OVLequal}
\end{equation}
This expression also holds when $\mu_1(\mathbf{z})=\mu_2(\mathbf{z})$, in which case $OVL(\mathbf{z})=1$.
\newline

Let $\{ (\mathbf{z}_{1r}, X_{1r}): r = 1, \ldots, n_1 \}$ and $\{ (\mathbf{z}_{2s}, X_{2s}): s = 1, \ldots, n_2 \}$ be independent random samples from Group 1 and Group 2, respectively, under models~\eqref{eq:LM1} and~\eqref{eq:LM2}, where $\mathbf{z}_{1r} = (1,z_{r1;1},z_{r2;1},\ldots,z_{rp;1})$
and $\mathbf{z}_{2s} = (1,z_{s1;2},z_{s2;2},\ldots,z_{sp;2})$.  For $i = 1, 2$, define:

\[\widetilde{\mathbf{X}}_i = (X_{i1}, \ldots, X_{in_i})^T, \quad \bm{\beta}_i =(\beta_{i0}, \beta_{i1}, \ldots, \beta_{ip}),\]

\begin{align*}
\widetilde{\mathbf{Z}}_i &=
\begin{bmatrix}
  1 & z_{11;i} & z_{12;i} & \ldots & z_{1p;i} \\
  1 & z_{21;i} & z_{22;i} & \ldots & z_{2p;i} \\
  \vdots & \vdots & \vdots & \ddots & \vdots \\
  1 & z_{n_i1;i} & z_{n_i2;i} & \ldots & z_{n_ip;i}
\end{bmatrix}.
\end{align*}

The regression coefficients can be estimated using ordinary least squares (OLS):
\[\widehat{\bm{\beta}}_i^T =(\widetilde{\mathbf{Z}}_i^T\widetilde{\mathbf{Z}}_i)^{-1}\widetilde{\mathbf{Z}}_i^T\widetilde{\mathbf{X}}_i,\quad i = 1, 2.\]

Consequently, the conditional means and the error standard deviations can be estimated as:
\[\widehat{\mu}_i(\mathbf{z})= \mathbf{z}\widehat{\bm{\beta}}_i^T, \quad i = 1, 2.\]

\[\widehat{\sigma}_i = \sqrt{\frac{
(\widetilde{\mathbf{X}}_i - \widetilde{\mathbf{Z}}_i\widehat{\bm{\beta}}_i^T)^T(\widetilde{\mathbf{X}}_i - \widetilde{\mathbf{Z}}_i\widehat{\bm{\beta}}_i^T)}{n_i-p-1}}
=\sqrt{\frac{\widetilde{\mathbf{X}}_i^T\widetilde{\mathbf{X}}_i -\widetilde{\mathbf{X}}_i^T\widetilde{\mathbf{Z}}_i\widehat{\bm{\beta}}_i^T}{n_i-p-1}},\quad i = 1, 2.\]

Now, substituting \(\widehat{\mu}_1(\mathbf{z}), \widehat{\sigma}_1, \widehat{\mu}_2(\mathbf{z}), \widehat{\sigma}_2\) into equations~\eqref{eq:OVLz1} (or into equation~\eqref{eq:OVLequal} in the equal-variance case), we obtain an estimator of OVL($\mathbf{z}$) denoted it by \(\widehat{OVL}(\mathbf{z})\).

\section{Generalized confidence intervals}
The concept of generalized confidence intervals was introduced by Tsui and Weerahandi (1989) and Weerahandi (1993), and has been applied to another summary measures of the ROC curve in presence of covariates, such as the Youden index and its optimal cut-off point as studied by Hu et al. (2021). However, this approach has not been directly explored for the  the covariate-specific overlap coefficient. First, we will establish the definition of a generalized pivotal quantity.
\newline

Let $W$ be a random variable with a distribution that depends on $(\theta, \delta)$, where $\theta$ is the parameter of interest and $\delta$ is a nuisance parameter. Let $w$ be an observed value of $W$. A function of $W$, $w$, $\theta$, and $\delta$, denoted as $R(W, w, \theta, \delta)$, is a generalized pivotal quantity (GPQ) if it satisfies the following two conditions:
\begin{enumerate}
    \item $R(W, w, \theta, \delta)$ has a distribution that is free from all unknown parameters.
    \item The observed value of $R(W, w, \theta, \delta)$ at $W = w$ does not depend on the nuisance parameter $\delta$.
\end{enumerate}

Following the methodology proposed by Hu et al. (2021), we first derive GPQs for the quantities \(\mu_1(\mathbf{z})\), \(\mu_2(\mathbf{z})\), \(\sigma_1\), and \(\sigma_2\), at a given covariate vector including an intercept $\mathbf{z}\in\mathbb{R}^{p+1}$. Conditional on the design matrix $\widetilde{\mathbf{Z}}_i$, the OLS estimator \(\widehat{\bm{\beta}}_i^T\) follows a multivariate normal distribution. Consequently, by standard properties of the normal distribution, we have:
\[\mathbf{z} \widehat{\bm{\beta}}_i^T \sim \mathcal{N}(\mathbf{z} \bm{\beta}_i^T, \sigma_i^2 V_i), \quad i = 1, 2,\]
with
\[\text{Var}(\mathbf{z} \widehat{\bm{\beta}}_i^T) = \mathbf{z} \, \text{Var}(\widehat{\bm{\beta}}_i^T) \, \mathbf{z}^T = \sigma_i^2 \mathbf{z} (\widetilde{\mathbf{Z}}_i^T \widetilde{\mathbf{Z}}_i)^{-1} \mathbf{z}^T = \sigma_i^2 V_i  \quad i = 1, 2.\]
\newline

For \(i=1,2\), let \(m_i(\mathbf{z})\) and \(s_i\) denote the observed values of \(\widehat{\mu}_i(\mathbf{z})\) and \(\widehat{\sigma}_i\), respectively. A GPQ for \(\mu_i(\mathbf{z}) = \mathbf{z}\bm{\beta}_i^T\) is then defined as
\[
R_{\mu_i(\mathbf{z})} =  m_i(\mathbf{z}) - (\mathbf{z} \widehat{\bm{\beta}}_i^T - \mathbf{z} \bm{\beta}_i^T) \left( \frac{s_i}{\widehat{\sigma}_i} \right).
\]

The distribution of \(R_{\mu_i(\mathbf{z})}\) can be obtained as

\begin{align*}
R_{\mu_i(\mathbf{z})} 
= m_i(\mathbf{z}) - (\mathbf{z} \widehat{\bm{\beta}}_i^T - \mathbf{z} \bm{\beta}_i^T) \left( \frac{s_i}{\widehat{\sigma}_i} \right) 
= m_i(\mathbf{z}) - \frac{\mathbf{z} \widehat{\bm{\beta}}_i^T - \mathbf{z} \bm{\beta}_i^T}{\sigma_i \sqrt{V_i}} \cdot \sigma_i \sqrt{V_i} \cdot \frac{s_i}{\widehat{\sigma}_i} 
= m_i(\mathbf{z})  - \mathcal{Z}_i \cdot \sigma_i \sqrt{V_i} \cdot \frac{s_i}{\widehat{\sigma}_i} \\
\end{align*}
\begin{align*}
= m_i(\mathbf{z}) - \frac{\mathcal{Z}_i}{\sqrt\frac{{\widehat{\sigma}_i^2}}{\sigma_i^2}} \cdot \sqrt{V_i} \cdot s_i=  m_i(\mathbf{z}) - \frac{\mathcal{Z}_i}{\sqrt{\frac{(n_i - p - 1)\widehat{\sigma}_i ^2}{(n_i - p - 1) \sigma_i^2}}} \cdot \sqrt{V_i} \cdot s_i= m_i(\mathbf{z}) - \frac{\mathcal{Z}_i}{\sqrt{\frac{\chi^2_{n_i - p - 1}}{n_i - p -1}}} \cdot \sqrt{V_i} \cdot s_i
\end{align*}
\begin{align}
= m_i(\mathbf{z})  - T_{n_i - p - 1} \cdot \sqrt{V_i} \cdot s_i.
\label{eq:medlineal1z}
\end{align}

Since, for \(i = 1, 2\), \(\mathcal{Z}_i\sim\mathcal{N}(0,1)\) and \(\frac{(n_i-p-1)\widehat{\sigma}_i^2}{\sigma_i^2}\sim\chi^2_{n_i-p-1}\) are independent, it follows that \(\frac{\mathcal{Z}_i}{\sqrt{\frac{\chi^2_{n_i-p-1}}{n_i-p-1}}}\sim T_{n_i-p-1}\), where \(T_{n_i-p-1}\) denotes a Student's \(t\)-distributed random variable with \(n_i-p-1\) degrees of freedom. The distribution of \(R_{\mu_i(\mathbf z)}\) is free from unknown parameters, and its observed value equals \(\mathbf z\bm{\beta}_i^T\). Therefore, \(R_{\mu_i(\mathbf z)}\) is a generalized pivotal quantity for \(\mu_i(\mathbf z)=\mathbf z\bm{\beta}_i^T\).
\newline

For the variance components \(\sigma_1^2\) and \(\sigma_2^2\), we employ the following generalized pivotal quantities:
\begin{align}
R_{\sigma_i^2} &= \sigma_i^2 \cdot \frac{s_i^2}{\widehat{\sigma}_i^2} = \frac{1}{\frac{(n_i - p - 1)\widehat{\sigma}_i^2}{\sigma_i^2}} \cdot (n_i - p - 1) s_i^2 = \frac{1}{\chi^2_{n_i - p - 1}} \cdot (n_i - p - 1) s_i^2, \quad i = 1, 2.
\label{eq:var1zgpq}
\end{align}

Under the normal linear regression models \eqref{eq:LM1} and \eqref{eq:LM2}, conditional on the observed design matrices, the distributions of \(R_{\sigma_i^2}\), for \(i=1,2\), are free from unknown parameters. Moreover, their observed values equal \(\sigma_i^2\), so, both quantities are GPQs for the corresponding variance parameters. It is natural to define
\begin{align}
R_{\sigma_i} = \sqrt{R_{\sigma_i^2}}
\label{eq:desvzgpq}
\end{align}
as a GPQ for \(\sigma_i\), \(i = 1, 2\).
\newline

By replacing \(\mu_i(\mathbf{z})\), \(\sigma_i\), and \(\sigma_i^2\) with their corresponding GPQs, \(R_{\mu_i(\mathbf{z})}\), \(R_{\sigma_i}\), and \(R_{\sigma_i^2}\), for \(i=1,2\), in equation~\eqref{eq:OVLz1} (or the corresponding expression for equal variances given by equation~\eqref{eq:OVLequal}), we obtain a GPQ for \(\text{OVL}(\mathbf{z})\), denoted by \(R_{\text{OVL}(\mathbf{z})}\). Indeed, the distribution of \(R_{\text{OVL}(\mathbf{z})}\) is free from unknown parameters, since it depends only on the Student's \(t\) and chi-squared random variables used to construct the GPQs for \(\mu_i(\mathbf{z})\), \(\sigma_i^2\), and \(\sigma_i\), \(i=1,2\). Furthermore, its observed value equals \(\text{OVL}(\mathbf{z})\).
\newline

Using Monte Carlo simulation, the distribution of \(R_{\text{OVL}(\mathbf{z})}\) can be approximated, and a \(100(1 - \alpha)\%\) generalized confidence interval can be obtained using the algorithm described below.

\begin{enumerate}
    \item Estimate \(\widehat{\bm{\beta}}_1\) and \(\widehat{\bm{\beta}}_2\) via OLS.

    \item Compute \(s_i\), for \(i=1,2\).

For a given \(\mathbf{z} \in \mathbb{R}^{p+1}\):

    \item Compute \(m_i(\mathbf{z})\) and \(V_i\), for \(i=1,2\).

    \item For \(m=1,\ldots,M\), repeat the following steps:
    \begin{enumerate}
        \item Independently generate \(\mathcal{Z}_i\sim\mathcal{N}(0,1)\), for \(i=1,2\).

        \item Independently generate \(\chi^2_{n_i-p-1}\) random variables, for \(i=1,2\), independently of the standard normal variables generated in step (a).

        \item Compute \(R_{\mu_i(\mathbf{z})}\), \(R_{\sigma_i^2}\), and \(R_{\sigma_i}\),
        according to equations~\eqref{eq:medlineal1z}-\eqref{eq:desvzgpq}, using
        \(T_{n_i-p-1}=\frac{\mathcal{Z}_i}{\sqrt{\chi^2_{n_i-p-1}/(n_i-p-1)}}\).

        \item Obtain \(R_{\text{OVL}(\mathbf{z})}\) by substituting the corresponding GPQs into
        equation~\eqref{eq:OVLz1}. Since the simulated standard deviations may not satisfy \(R_{\sigma_1}<R_{\sigma_2}\), interchange the group labels whenever necessary to ensure this ordering. If the absolute difference between the simulated variance GPQs is smaller than a prespecified tolerance, the equal-variance expression given by equation~\eqref{eq:OVLequal} is used as a numerical approximation.
    \end{enumerate}

    \item Compute the empirical \(\alpha/2\) and \(1-\alpha/2\) quantiles from the set \(\{R_{\text{OVL},1}(\mathbf{z}),\ldots,R_{\text{OVL},M}(\mathbf{z})\}\). The resulting interval
    \[
    \left(R_{\text{OVL}(\mathbf{z}),\alpha/2},\,R_{\text{OVL}(\mathbf{z}),1-\alpha/2}\right)
    \]
    defines a \(100(1-\alpha)\%\) generalized confidence interval for \(\text{OVL}(\mathbf{z})\).
\end{enumerate}

\vspace{1em}

\noindent The assumption that the variables $X_1|(\mathbf{Z}=\mathbf{z})$ and $X_2|(\mathbf{Z}=\mathbf{z})$ are normally distributed can be quite restrictive leading to false results when it is significantly violated. A popular way of extending this approach is to assume that a common, usually unknown, monotone transformation results in both distributions being normal. The Box-Cox transformation (Box and Cox, 1964) is defined as:
\begin{equation}
X_i^{(\lambda)} \mid (\mathbf{Z} = \mathbf{z}) =
\begin{cases}
    \log(X_i \mid (\mathbf{Z} = \mathbf{z})) & \text{if } \lambda = 0 \\
    \dfrac{(X_i^\lambda \mid (\mathbf{Z} = \mathbf{z})) - 1}{\lambda} & \text{if } \lambda \neq 0
\end{cases}
\quad,\quad i = 1, 2.
\label{eq: Box-Cox}
\end{equation}

\noindent The maximum likelihood estimator of the transformation parameter \(\lambda\) can be obtained by maximizing the log-likelihood function:
\begin{equation*}
\begin{split}
L(\lambda) = & -\frac{n_1}{2} \log\left(\frac{\sum_{i = 1}^{n_1} \left(X_{1i}^{(\lambda)} - \mathbf{z}_{1i} \cdot (\widehat{\bm{\beta}}_1^{(\lambda)})^T \right)^2}{n_1} \right) 
- \frac{n_2}{2} \log\left(\frac{\sum_{j = 1}^{n_2} \left(X_{2j}^{(\lambda)} - \mathbf{z}_{2j} \cdot (\widehat{\bm{\beta}}_2^{(\lambda)})^T \right)^2}{n_2} \right) \\
& + (\lambda - 1) \left( \sum_{i = 1}^{n_1} \log(X_{1i}) + \sum_{j = 1}^{n_2} \log(X_{2j}) \right) + k,
\end{split}
\label{eq:logverosimilitud}
\end{equation*}
\noindent where \(k\) is a constant. Once the Box–Cox parameter is estimated, the transformation in \eqref{eq: Box-Cox} is applied to both samples, and the previously described algorithm is applied to the transformed data.

\section{Simulation results}

In this section, we evaluate the performance of the GPQ-based method under the assumption of a linear relationship between the covariate and the biomarkers. To further enrich the simulation study, we additionally consider several bootstrap methods proposed in Pardo et al. (2025), which also model the relationship between the covariate and the biomarkers through linear regression models. Specifically, we include the following approaches: (i) a method based on the normal approximation of the confidence interval, where the variance is estimated via bootstrap, combined with the logit-scale and the Box–Cox transformation; (ii) the percentile bootstrap approach applied after the Box–Cox transformation; and (iii) the bias-corrected bootstrap approach applied after the Box–Cox transformation. These methods were denoted as L-BC-AN, BC-PB, and BC-bias, respectively.
\newline

\noindent Six scenarios were considered: two in which \(X_1|\mathbf{Z}\) and \(X_2|\mathbf{Z}\) follow Normal distributions, two where both variables follow PowerNormal distributions, and another two where \(X_1|\mathbf{Z}\) and \(X_2|\mathbf{Z}\) follow LogNormal distributions. In all scenarios, we have considered one covariate that follows a \( U(0,1) \) distribution. Table~\ref{tab:ParametersSI} summarizes the parameters of the distributions considered. The true OVL value for each scenario, for a fixed covariate value, can be found in the Supplementary~Material (Table 1). These scenarios were first considered in Pardo et al. (2025).
\newline

\begin{table}[H]
\centering
\renewcommand{\arraystretch}{1.2} 
\begin{tabular}{c|c|c|c}
\hline
\textbf{Scenario} & \textbf{Base Model (Normal)} & \textbf{LogNormal} & \textbf{PowerNormal} \\
\hline
1 &
\begin{tabular}[c]{@{}l@{}}
$X_1|\mathbf{Z} \sim  \mathcal{N}(6.5 + 2Z,\, 0.09) $ \\ 
$X_2|\mathbf{Z} \sim  \mathcal{N}(6.873 + 1.627Z,\, 0.125)$ 
\end{tabular}
&
\begin{tabular}[c]{@{}l@{}}
$\exp(X_1|\mathbf{Z})$ \\
$\exp(X_2|\mathbf{Z})$
\end{tabular}
&
\begin{tabular}[c]{@{}l@{}}
$(X_1|\mathbf{Z})^{-1/3}$ \\
$(X_2|\mathbf{Z})^{-1/3}$
\end{tabular}
\\
\hline
2 &
\begin{tabular}[c]{@{}l@{}}
$X_1|\mathbf{Z} \sim  \mathcal{N}(6.5 + 2Z,\, 0.09)$ \\ 
$X_2|\mathbf{Z} \sim  \mathcal{N}(6.873 + Z,\, 0.125)$
\end{tabular}
&
\begin{tabular}[c]{@{}l@{}}
$\exp(X_1|\mathbf{Z})$ \\
$\exp(X_2|\mathbf{Z})$
\end{tabular}
&
\begin{tabular}[c]{@{}l@{}}
$(X_1|\mathbf{Z})^{-1/3}$ \\
$(X_2|\mathbf{Z})^{-1/3}$
\end{tabular}
\\
\hline
\end{tabular}
\caption{Base Normal models and corresponding LogNormal and PowerNormal transformations.}
\label{tab:ParametersSI}
\end{table}

\begin{figure}[ht]
    \centering
    \begin{subfigure}{0.49\textwidth}
         \includegraphics[width=\linewidth, height=5cm, keepaspectratio]{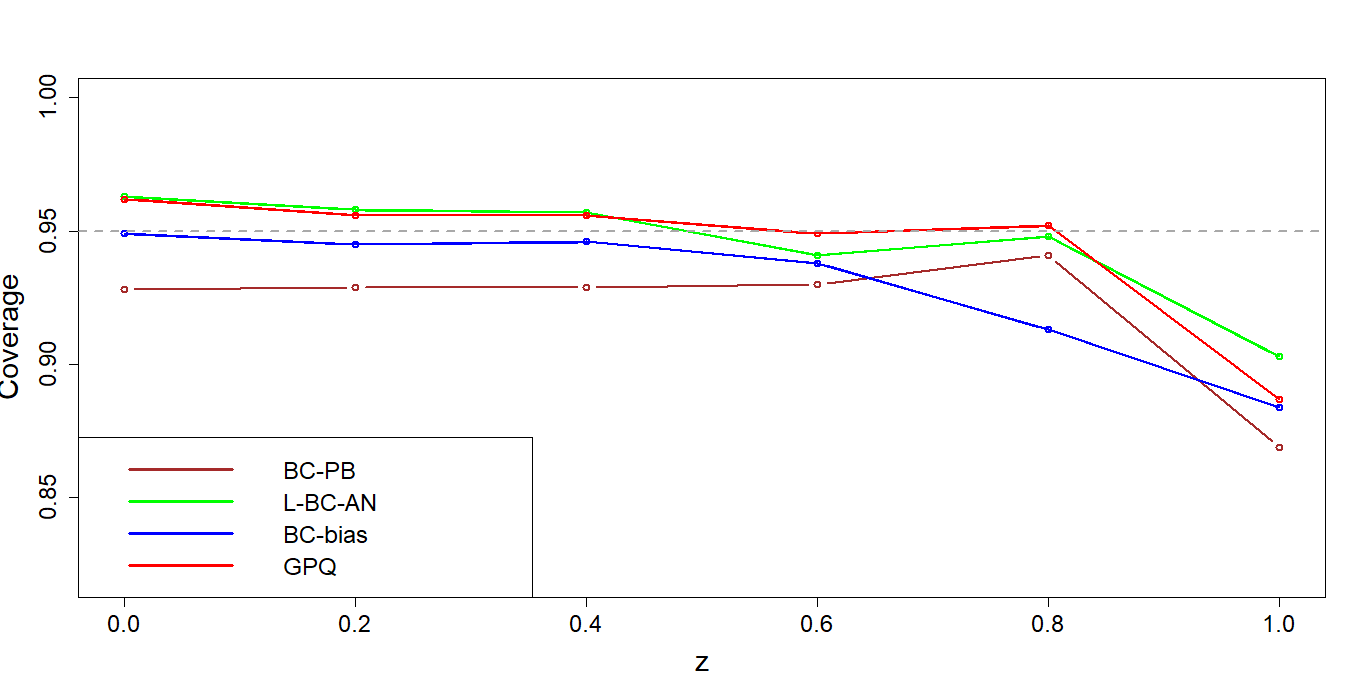}
      
        \caption*{(50,50)}
    \end{subfigure}
    \begin{subfigure}{0.49\textwidth}
    \includegraphics[width=\linewidth, height=5cm, keepaspectratio]{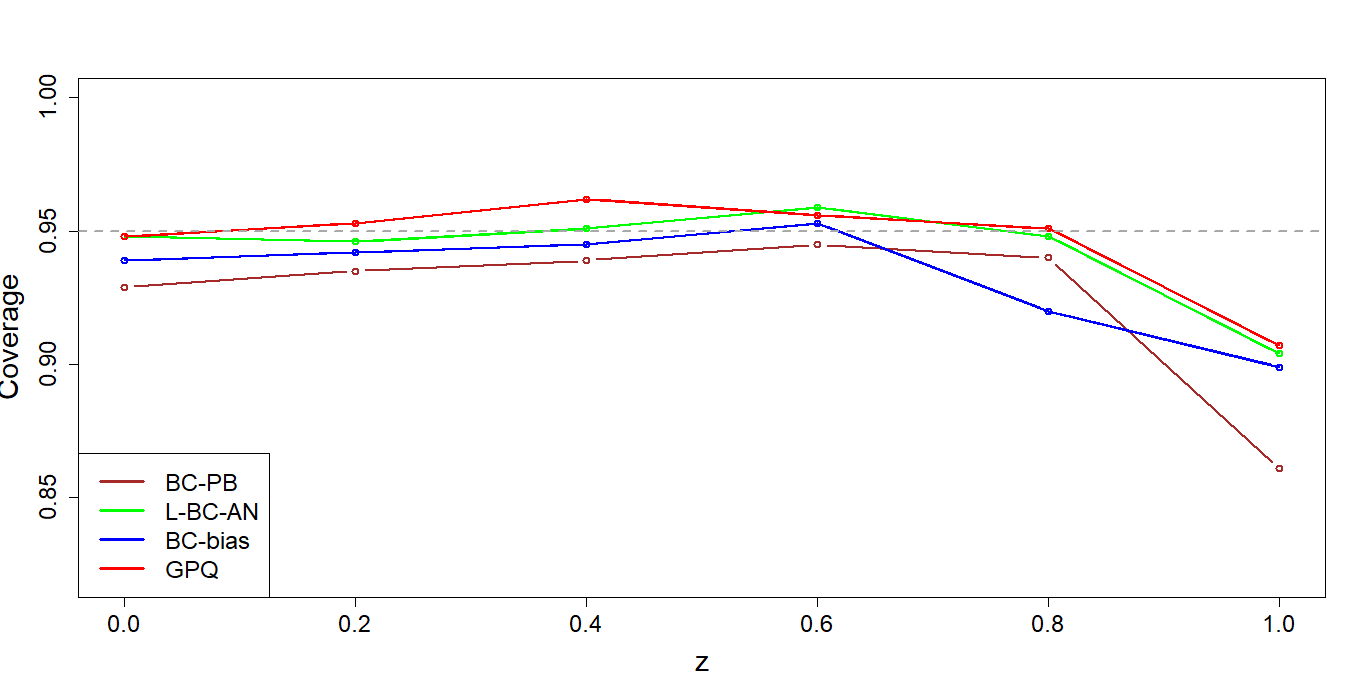}
        \caption*{(50,100)}
    \end{subfigure}
    \begin{subfigure}{0.49\textwidth}
        \includegraphics[width=\linewidth, height=5cm, keepaspectratio]{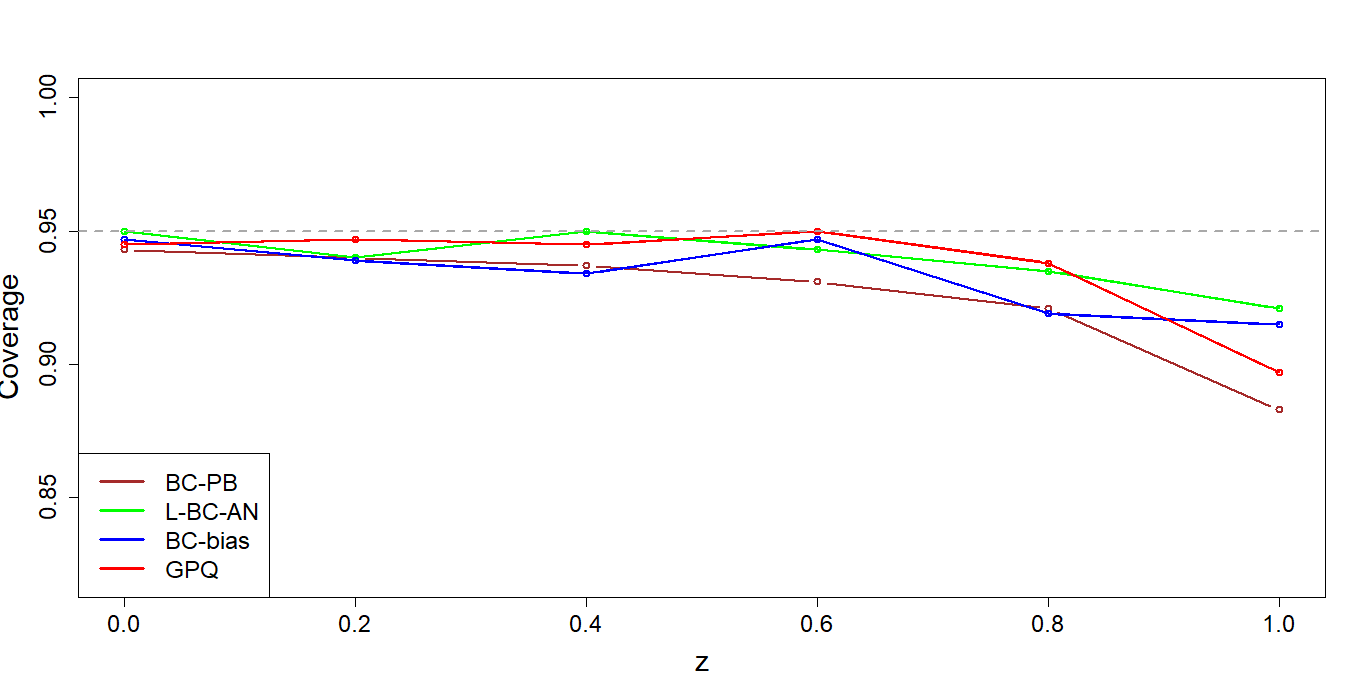}
        \caption*{(100,100)}
    \end{subfigure}
    \begin{subfigure}{0.49\textwidth}
       \includegraphics[width=\linewidth, height=5cm, keepaspectratio]{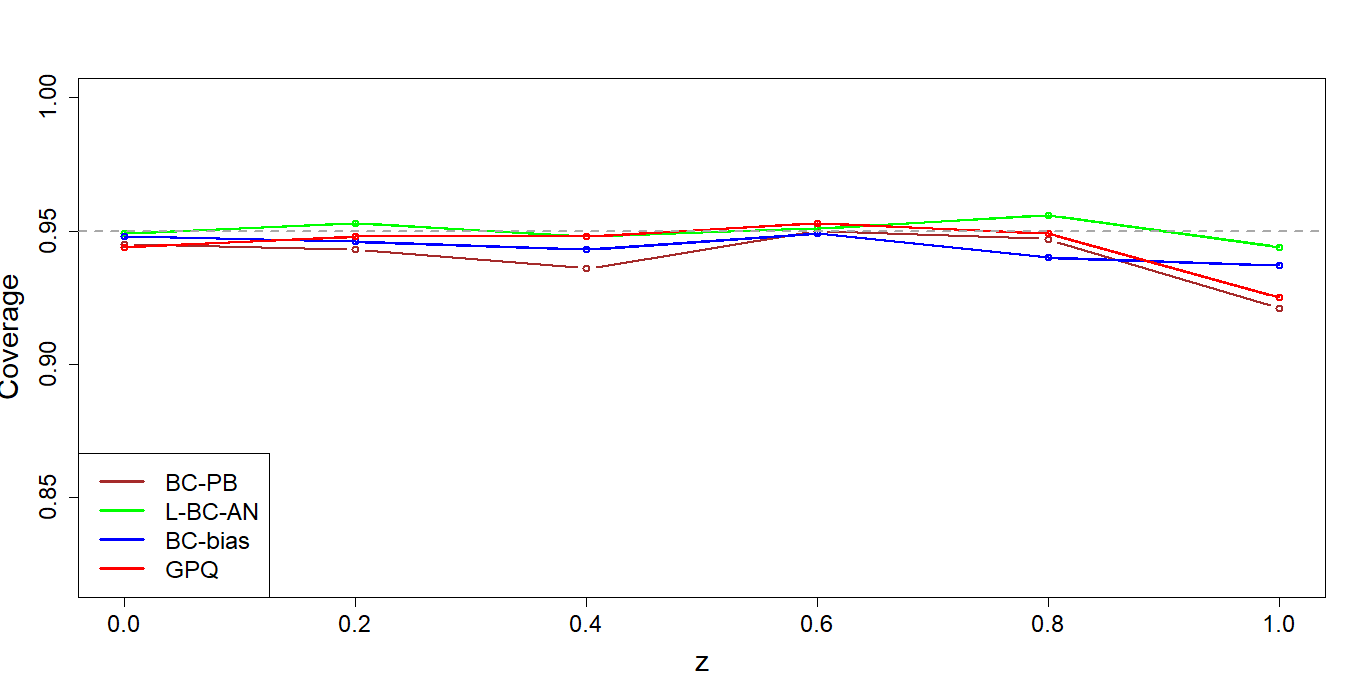}
        \caption*{(200,200)}
    \end{subfigure}
     \caption{Coverages of GPQ and bootstrap methods for the Normal case, scenario 1}
    \label{fig:coberturasnormal1}
\end{figure}

\begin{figure}[ht]
    \centering
    \begin{subfigure}{0.49\textwidth}
       \includegraphics[width=\linewidth, height=5cm, keepaspectratio]{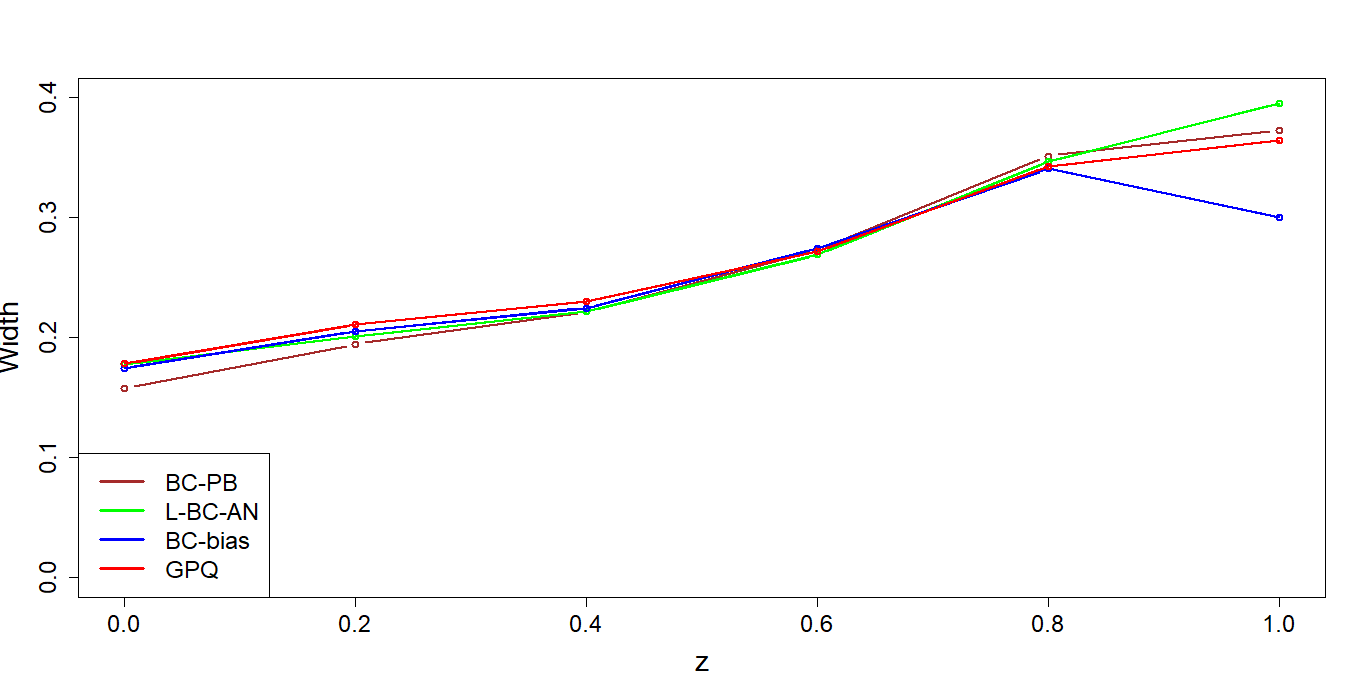}
        \caption*{(50,50)}
    \end{subfigure}
    \begin{subfigure}{0.49\textwidth}
         \includegraphics[width=\linewidth, height=5cm, keepaspectratio]{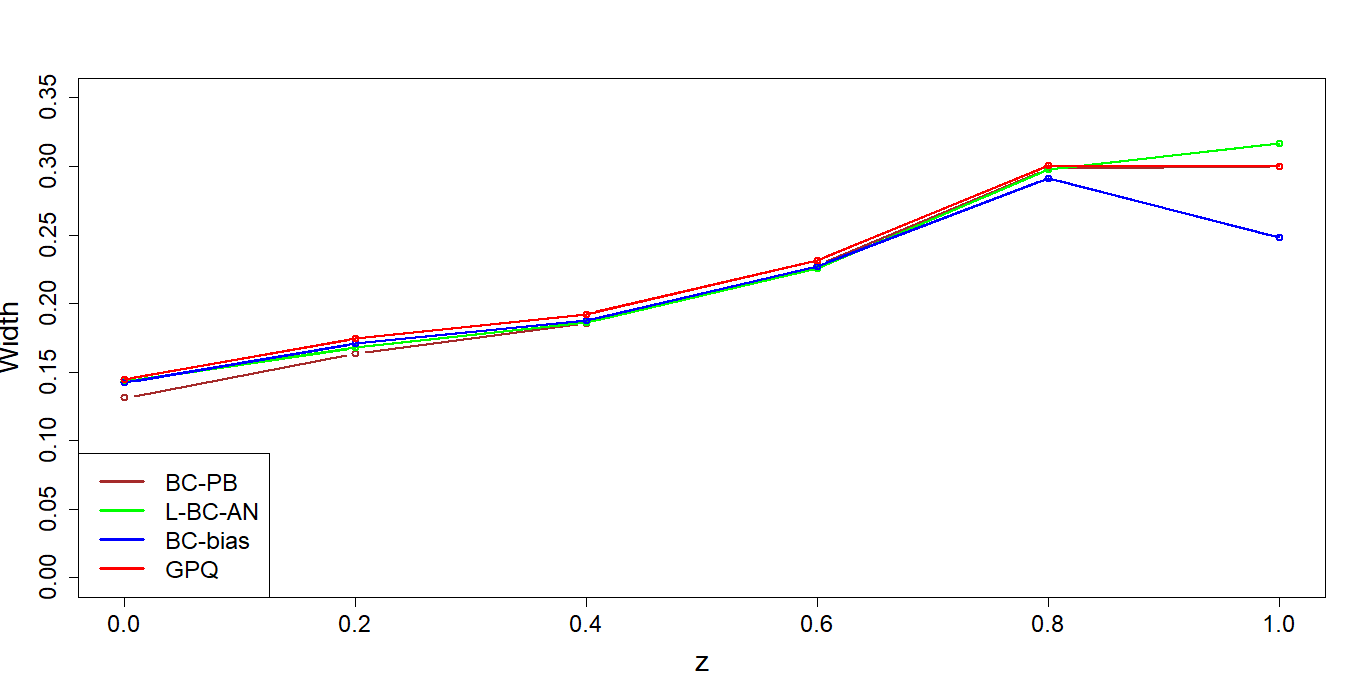}
        \caption*{(50,100)}
    \end{subfigure}
    \begin{subfigure}{0.49\textwidth}
        \includegraphics[width=\linewidth, height=5cm, keepaspectratio]{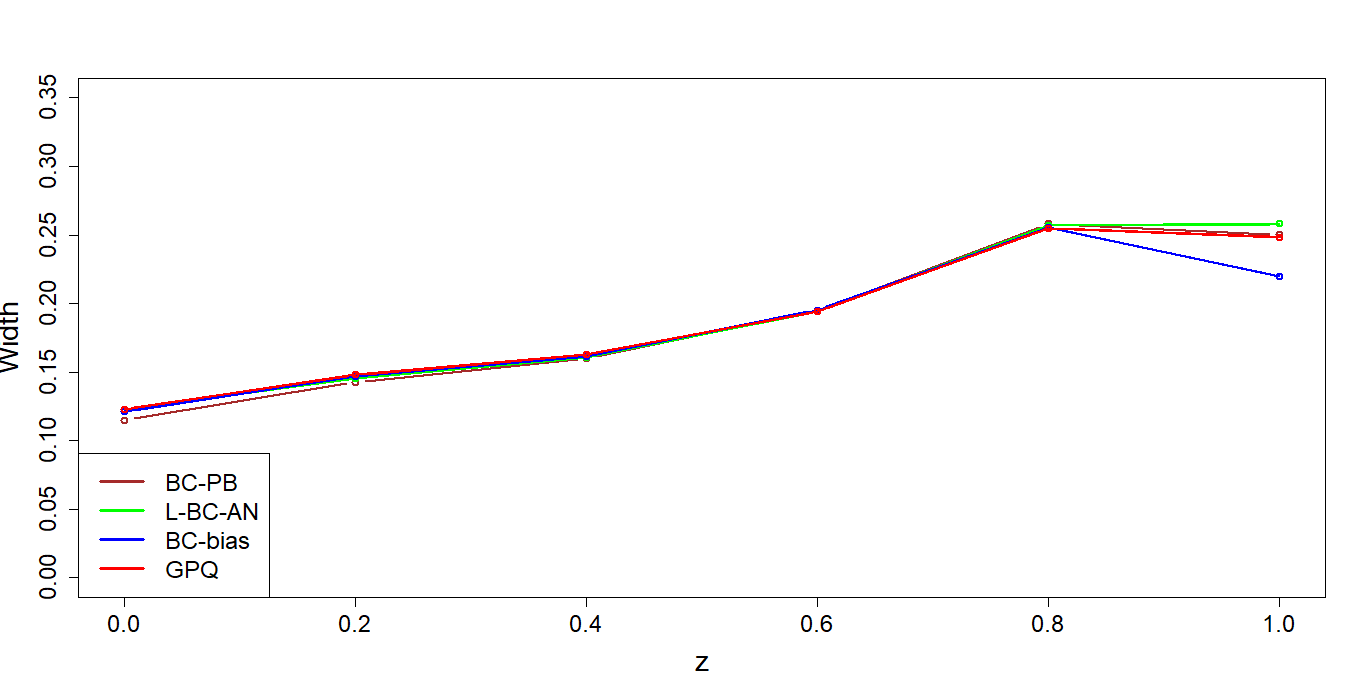}
        \caption*{(100,100)}
    \end{subfigure}
    \begin{subfigure}{0.49\textwidth}
        \includegraphics[width=\linewidth, height=5cm, keepaspectratio]{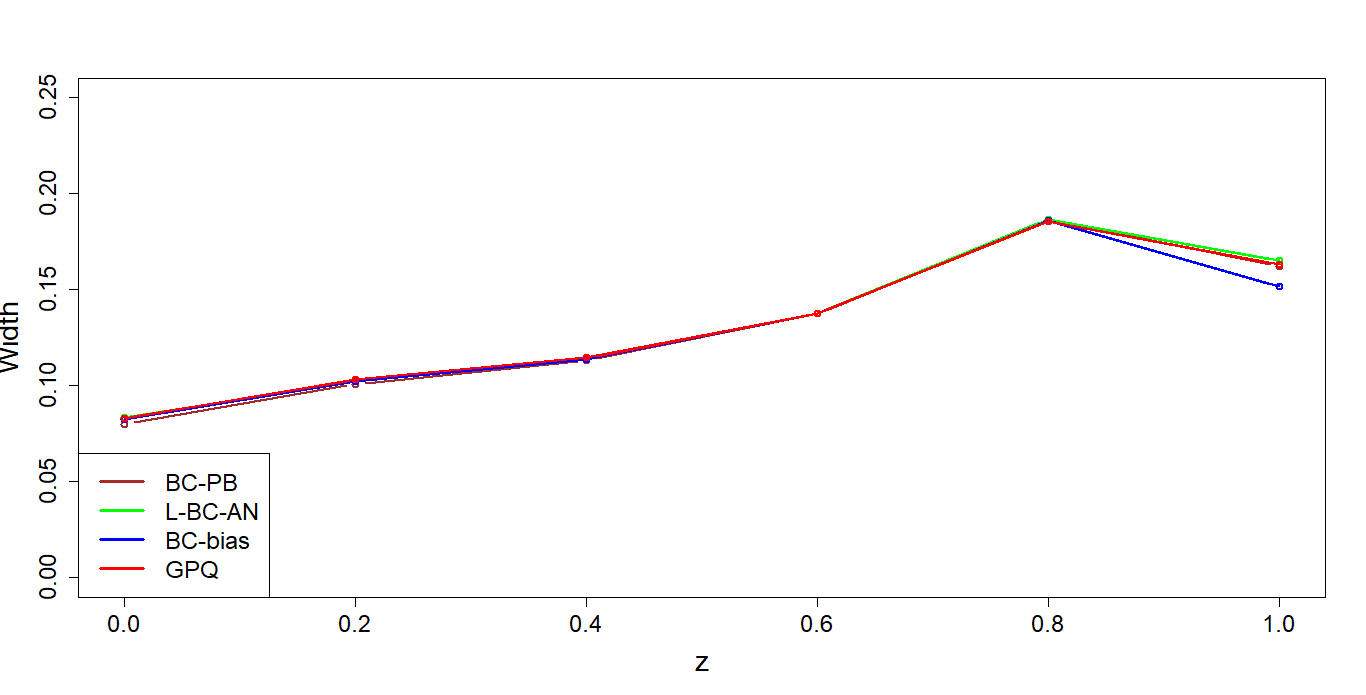}
        \caption*{(200,200)}
    \end{subfigure}
     \caption{Mean interval widths of GPQ and bootstrap methods for the Normal case, scenario 1}
     \label{fig:MILnormal1}
\end{figure}

Regarding the sample sizes, we considered balanced configurations \(n_1 = n_2 = 50, 100,\) and \(200\), as well as an unbalanced case with \(n_1 = 50\) and \(n_2 = 100\). We used $MC=1000$ Monte Carlo replications, $B=500$ bootstrap samples, and 
$M=500$ simulated values from the distribution of \(R_{\text{OVL}(\mathbf{z})}\) for the GPQ-based method. In all scenarios, 95\% confidence intervals were constructed, and the performance of the methods was evaluated through the computation of coverage proba\-bilities and mean interval widths. Figures~\ref{fig:coberturasnormal1} and~\ref{fig:MILnormal1} display, respectively, the coverages and the mean interval widths for the first Normal scenario across the covariate values. The results for the remaining scenarios can be found in the Supplementary Material (Figures~1-10).
\newline

\noindent Figure \ref{fig:coberturasnormal1} shows that, as the sample size increases, the coverage probabilities of all methods approach the nominal 95\% level, as expected. In general, the GPQ approach exhibits the most stable performance across different sample sizes, while the bootstrap-based methods show slightly more variability, particularly for small or unbalanced samples. Among the bootstrap methods, the bias-corrected version tends to achieve higher coverage, especially when the covariate approaches its boundary values. The method that exhibits the weakest performance is BC-PB. Nevertheless, all methods perform comparably for large samples, with only minor differences in coverage. Regarding the mean interval widths, Figure~\ref{fig:MILnormal1} shows that the values are very similar across all methods for each covariate value. Noticeable differences arise near $z = 1$, where the BC-bias method yields narrower intervals than the others, while the L-BC-AN method produces slightly wider ones. The conclusions for the second Normal scenario (Figures 1–2 in the Supplementary Material) are qualitatively similar, although in that case the differences among methods become more pronounced for intermediate values of OVL.
\newline

\noindent The results for the PowerNormal and LogNormal scenarios (Figures 3–10 in the Supplementary Material) are consistent with those obtained for the Normal case, as expected given the suitability of the Box–Cox transformation for these distributions. 
\newline

\noindent Another relevant aspect is that the GPQ method is computationally more efficient than the bootstrap algorithms, as they avoid the resampling process. For illustration, Table~\ref{tab:ratios_normal1} reports the relative computational cost of bootstrap algorithms compared with the GPQ-based method in a Normal scenario.

\begin{table}[H]
\centering
\begin{tabular}{rccc}
  \hline
\((n_1, n_2)\) & Ratio BC-PB/GPQ & Ratio L-BC-AN/GPQ & Ratio BC-bias/GPQ \\ 
  \hline
  (50,50)   & 9.50 & 9.40 & 9.37 \\
  (50,100)  & 9.73 & 9.73 & 9.57 \\
  (100,100) & 9.23 & 9.16 & 9.08 \\
  (200,200) & 9.34 & 9.00 & 9.04 \\
  \hline
\end{tabular}
\caption{Ratios of execution times for the Normal case, scenario 1}
\label{tab:ratios_normal1}
\end{table}

\section{Application to real data}
The practical performance of the GPQ-based method was further assessed using the \textit{Pima Indian Diabetes} dataset, originally released by the \textit{National Institute of Diabetes and Digestive and Kidney Diseases}. This database contains observations from women of Pima Indian ancestry aged between 21 and 81 years. It includes 768 individuals, of whom 500 are classified as non-diabetic (controls) and 268 as diabetic (cases). The dataset records nine variables: number of pregnancies, plasma glucose concentration measured by an oral glucose tolerance test, diastolic blood pressure, triceps skinfold thickness, serum insulin level, body mass index, diabetes pedigree function (which quantifies hereditary risk based on family history), age, and a binary outcome indicating disease status. This dataset was previously analyzed in Alcaraz-Peñalba et al.\ (2026), where 95\% confidence intervals were obtained using the most promising inference methods without accounting for the effect of covariates: the L-\(\delta\)-BC-\(\lambda\) method (Franco-Pereira et al., 2021) and the GPQ approach in the absence of covariates (Wang and Tian, 2017). The estimated OVL was 0.556, and the corresponding confidence intervals obtained using these two methods were (0.523, 0.633) and (0.516, 0.634), respectively, showing that the test measuring plasma glucose levels is a useful biomarker for distinguishing between diabetic and non-diabetic individuals.
\newline

Previous studies have suggested that individual characteristics may influence plasma glucose levels, and age has often been highlighted as a relevant clinical factor (Smith and Thompson, 1996). Motivated by this consideration, we examine how the discriminative performance of the oral glucose tolerance test varies across different age levels. Following Hu et al. (2021), five observations with zero glucose values were excluded from the analysis, as these values were considered measurement errors. Moreover, since this dataset was previously investigated by Pardo et al. (2025) in the context of constructing confidence intervals for covariate-specific OVL using bootstrap techniques, an additional goal of this section is to compare the inference provided by the proposed GPQ-based method with the bootstrap approaches considered in their study.
\newline

\begin{figure}[ht]
    \centering
    \begin{subfigure}{0.4\textwidth}
        \includegraphics[width=\linewidth, height=5cm]{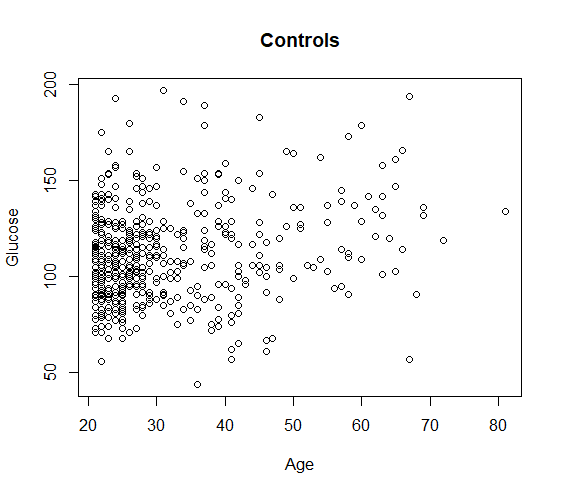}
    \end{subfigure}
    \begin{subfigure}{0.4\textwidth}
        \includegraphics[width=\linewidth, height=5cm]{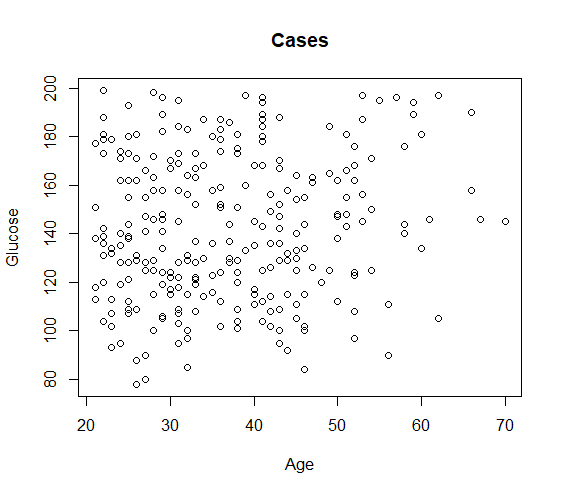}
    \end{subfigure}
   \caption{Scatter plots of plasma glucose level from the oral glucose tolerance test versus age}
    \label{fig:scatter}
\end{figure}

Figure~\ref{fig:scatter} indicates that the association between plasma glucose level and age is markedly nonlinear. Furthermore, in the control group, the variability of glucose levels appears to differ between individuals aged 20–30 and those in other age ranges, which may indicate heteroscedasticity in the residuals. To stabilize the variance of the regression errors, glucose levels were analyzed on a logarithmic scale for both groups, and models~(\ref{eq:LM1}) and (\ref{eq:LM2}) were fitted to describe the relationship between the log-transformed biomarker and age. The Breusch–Pagan (B–P) test for the model without transformation yielded a p-value of 0.003 for the control group. After applying the logarithmic transformation to the biomarker in both groups, the B–P test produced p-values of 0.011 for controls and 0.441 for cases, allowing us to assume homoscedasticity of the residuals at the 1\% significance level, but not at the 5\% significance level for the control group. The Kolmogorov–Smirnov (K–S) test yielded p-values of 0.9366 for the control residuals and 0.265 for the case residuals, indicating that normality of the errors can be assumed for both models. Although the logarithmic or Box–Cox transformation may help stabilize the residual variance, it does not necessarily ensure homoscedasticity or a linear relationship between the transformed glucose levels and age, which represents a potential limitation of the present analysis, as further discussed in the Concluding Remarks section.
\newline

To obtain pointwise estimates of the overlap coefficient for each age between 21 and 81 years, we employed the estimator \( \widehat{OVL}(\mathbf{z}) \). The 95\% confidence intervals were constructed using the L-BC-AN, BC-bias, and BC-PB methods proposed by Pardo et al. (2025), with \(B = 500\), as well as our proposed GPQ method, with \(M = 500\).
\newline

\begin{figure}[ht]
\centering
\includegraphics[width=0.9\linewidth]{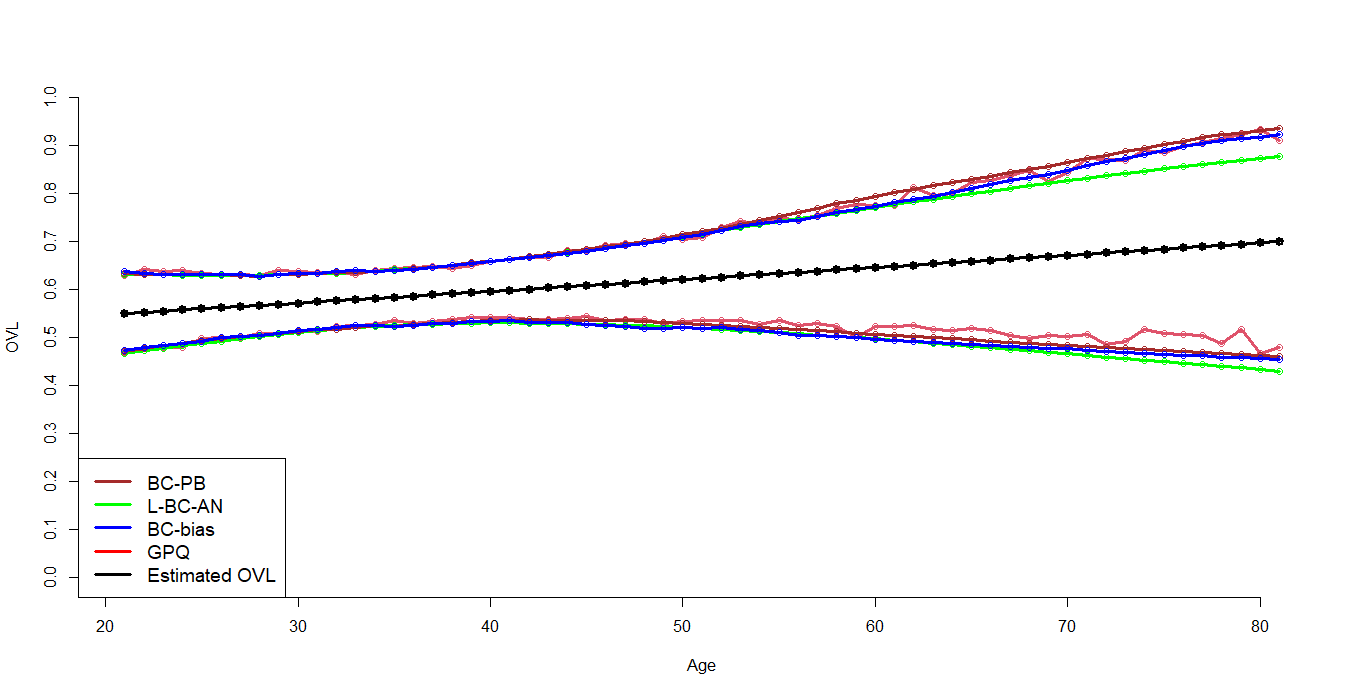}
\caption{OVL estimates adjusted for age.}
\label{fig:caso practico}
\end{figure}

Figure~\ref{fig:caso practico} shows that the discriminative ability of glucose levels to distinguish between diabetic and non-diabetic individuals decreases with age. The confidence intervals corresponding to each covariate value are generally similar across methods, except perhaps at more advanced ages. For older individuals, the confidence intervals tend to widen, indicating greater uncertainty regarding the true value of the overlap coefficient. It is worth noting that only two controls and one case were aged 70 years or older. Specifically, there was one control and one case aged 70–74, no observations aged 75–79, and only one control aged 80 or above. Thus, the estimates at advanced ages should be interpreted with caution, as they rely on extrapolation of the fitted regression model. Among the methods considered, L-BC-AN and GPQ tend to yield narrower intervals at older ages. Recall that the OVL estimate obtained without adjusting for age was 0.556. Thus, failing to account for this covariate would lead to an overestimation of the biomarker’s discriminative capacity at advanced ages. For instance, for individuals aged 81 years, the estimated OVL (approximately 0.70) would not even fall within the 95\% confidence intervals produced by the GPQ method in the unadjusted analysis, highlighting the substantial impact of age adjustment on the inference.

\section{Concluding Remarks}
The overlap coefficient is an easily interpretable measure that can be applied in many scientific fields where the main objective is to assess the similarity between probability distributions. One area in which this measure is particularly relevant is medical diagnosis, where evaluating the discriminatory ability of biomarkers and diagnostic tests is essential.
\newline

The development of inference methods for the covariate-specific OVL is crucial. Ignoring covariate effects may lead to overestimating or underestimating the discriminatory performance, ultimately resulting in misleading conclusions. A natural way to incorporate the effect of covariates on biomarkers is through regression modeling. In this work, we employed multiple linear regression to describe the relationship between the biomarker and the covariates, and we proposed a new inference method based on the construction of generalized pivotal quantities. The simulation study indicates that the GPQ-based approach tends to outperform some of the inference methods introduced by Pardo et al.\ (2025), particularly in small and unbalanced samples. Moreover, the generalized inference method stands out for its computational efficiency when compared with bootstrap-based procedures previously proposed in the literature.
\newline

Nevertheless, in practice, the assumptions underlying the linear relationship between the biomarker and the covariates, as well as those of the classical linear regression model, may not hold. Although Box–Cox transformations can provide additional flexibility, they may still be insufficient in certain situations. Future research will therefore focus on developing algorithms for constructing covariate-specific confidence intervals under heteroscedastic regression models, with or without assumptions regarding error normality. Additionally, we aim to compare the results obtained in the real-data application, where potential homoscedasticity and non-linear issues were observed, with those produced by more flexible regression models.
\newline

An \texttt{R} implementation of the covariate-specific GPQ method proposed in this work is available upon request from the first author and will be released on CRAN in the near future as part of the \texttt{OVL.CI} package.
\newline

\noindent \textbf{Acknowledgments}

This work was supported by grant PID2022-137050NB-I00 of the Spanish Ministry of Science and Innovation, funded by MCIN/AEI/10.13039/501100011033/ and by ERDF A way of making Europe. AAP was awarded a research assistant contract from the Community of Madrid (PEJ-2023-AI/SAL-GL-28227).

\newpage

\section*{Supplementary Material: Generalized confidence intervals for the covariate-specific overlap coefficient}
\setcounter{figure}{0}
\setcounter{table}{0}

\begin{table}[H]
\centering
\begin{adjustbox}{scale=1.1}
\subfloat[Scenario 1]{
\begin{tabular}{rr}
True OVL & $z$ \\
\hline
0.081 & 0  \\
0.162 & 0.2\\
0.292 & 0.4 \\
0.476 & 0.6 \\
0.701 & 0.8 \\
0.842 & 1 \\
\hline
\end{tabular}%
}
\quad 
\subfloat[Scenario 2]{ 
\begin{tabular}{rrrrr}
True OVL & $z$ \\
\hline
0.081 & 0  \\
0.412 & 0.2\\
0.820 & 0.4 \\
0.285 & 0.6 \\
0.046 & 0.8 \\
0.004 & 1 \\
\hline
\end{tabular}%
} 
\end{adjustbox}
\caption{True OVL for given covariate values}
\label{tab:ovlS1}
\end{table}

\begin{figure}[H]
    \centering
    \begin{subfigure}{0.49\textwidth}
         \includegraphics[width=\linewidth, height=5cm, keepaspectratio]{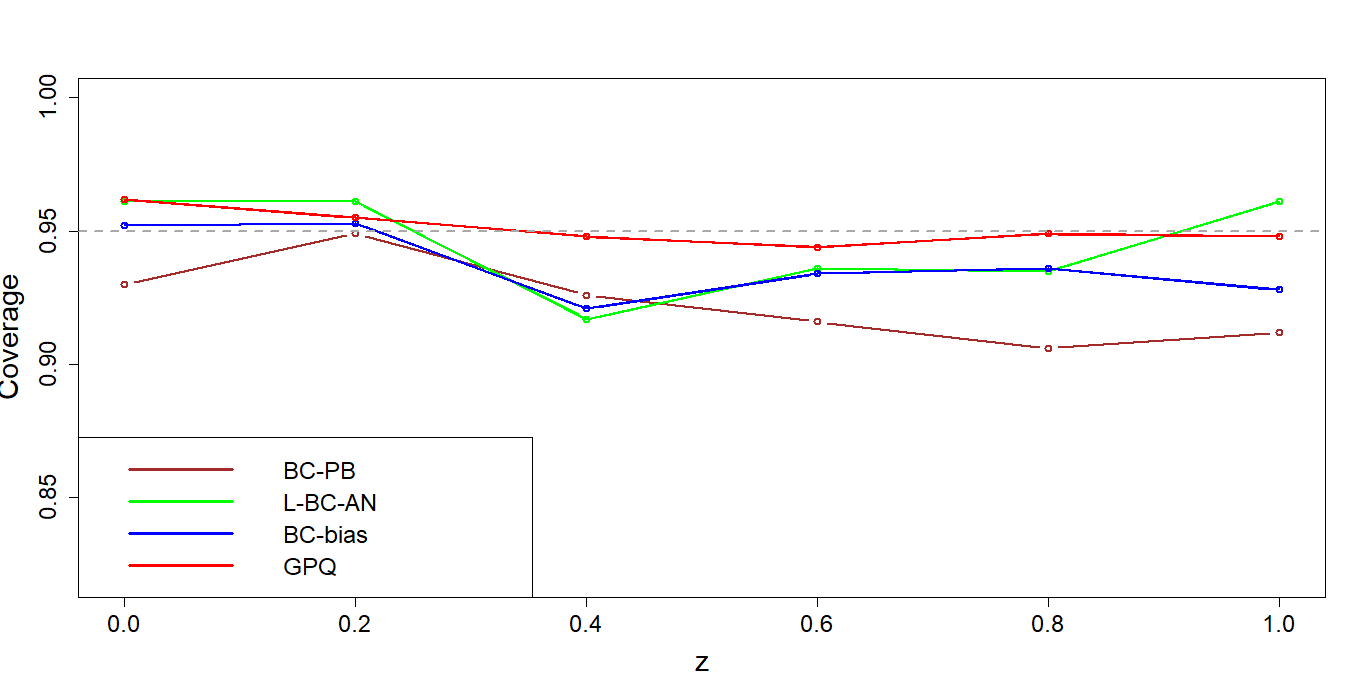}
      
        \caption*{(50,50)}
    \end{subfigure}
    \begin{subfigure}{0.49\textwidth}
    \includegraphics[width=\linewidth, height=5cm, keepaspectratio]{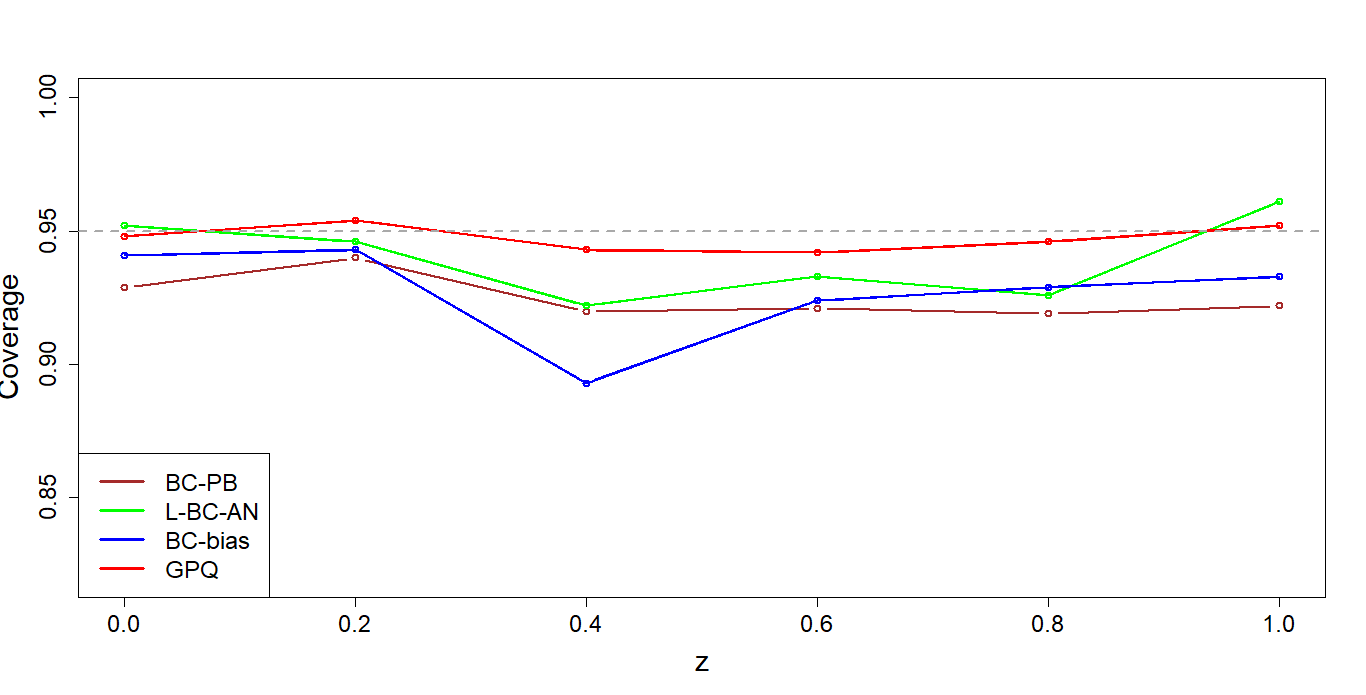}
        \caption*{(50,100)}
    \end{subfigure}
    \begin{subfigure}{0.49\textwidth}
        \includegraphics[width=\linewidth, height=5cm, keepaspectratio]{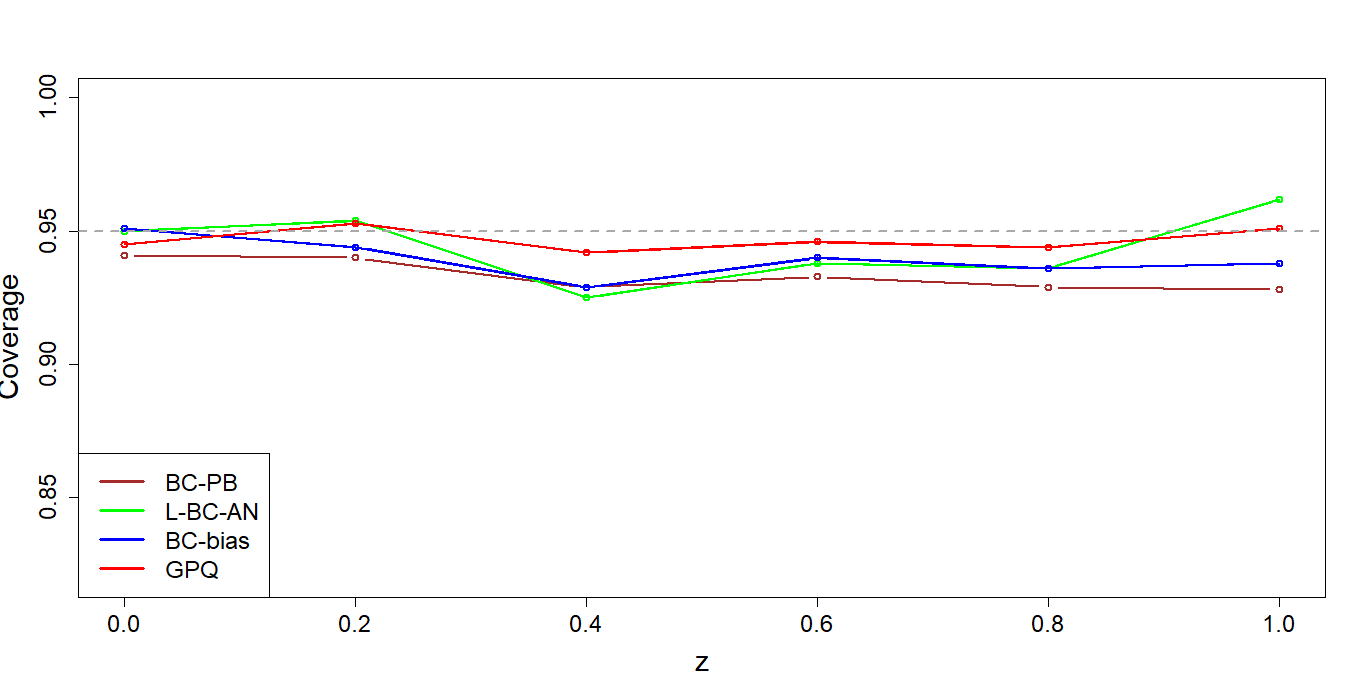}
        \caption*{(100,100)}
    \end{subfigure}
    \begin{subfigure}{0.49\textwidth}
       \includegraphics[width=\linewidth, height=5cm, keepaspectratio]{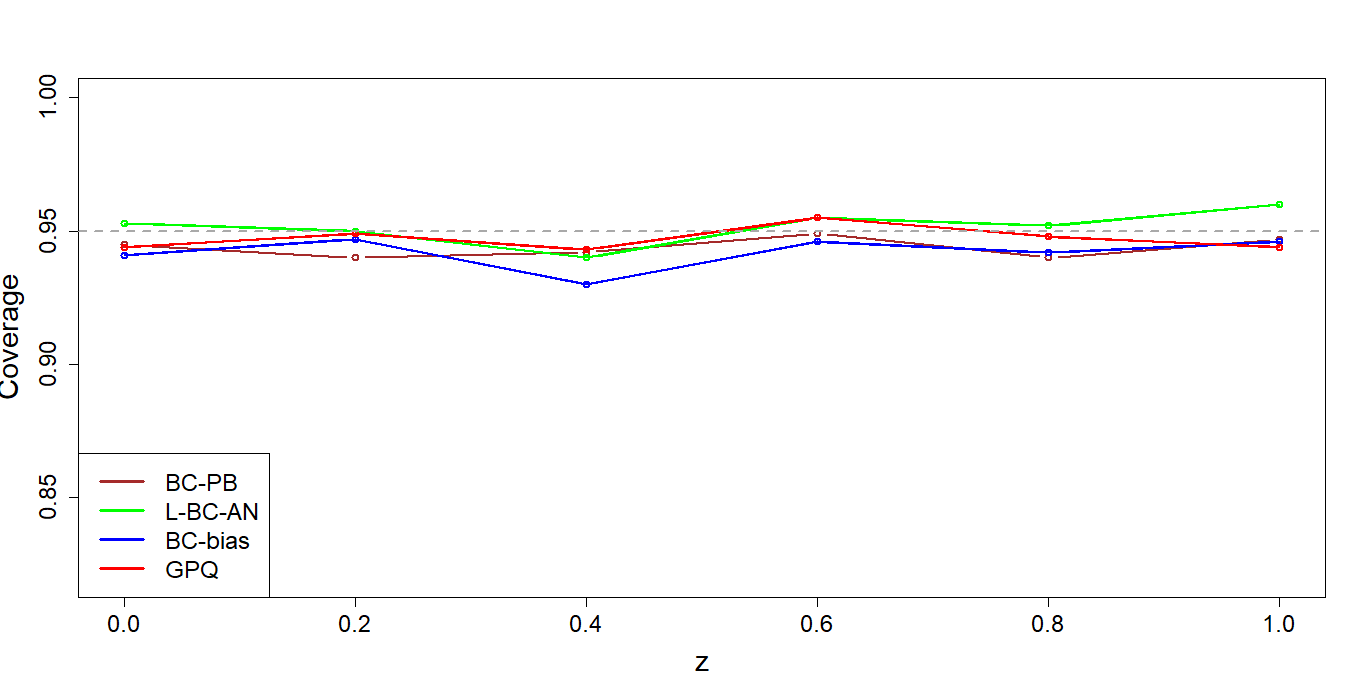}
        \caption*{(200,200)}
    \end{subfigure}
     \caption{Coverages of GPQ and bootstrap methods for the Normal case, scenario 2}
    \label{fig:coberturasnormal2}
\end{figure}

\begin{figure}
    \centering
    \begin{subfigure}{0.49\textwidth}
         \includegraphics[width=\linewidth, height=5cm, keepaspectratio]{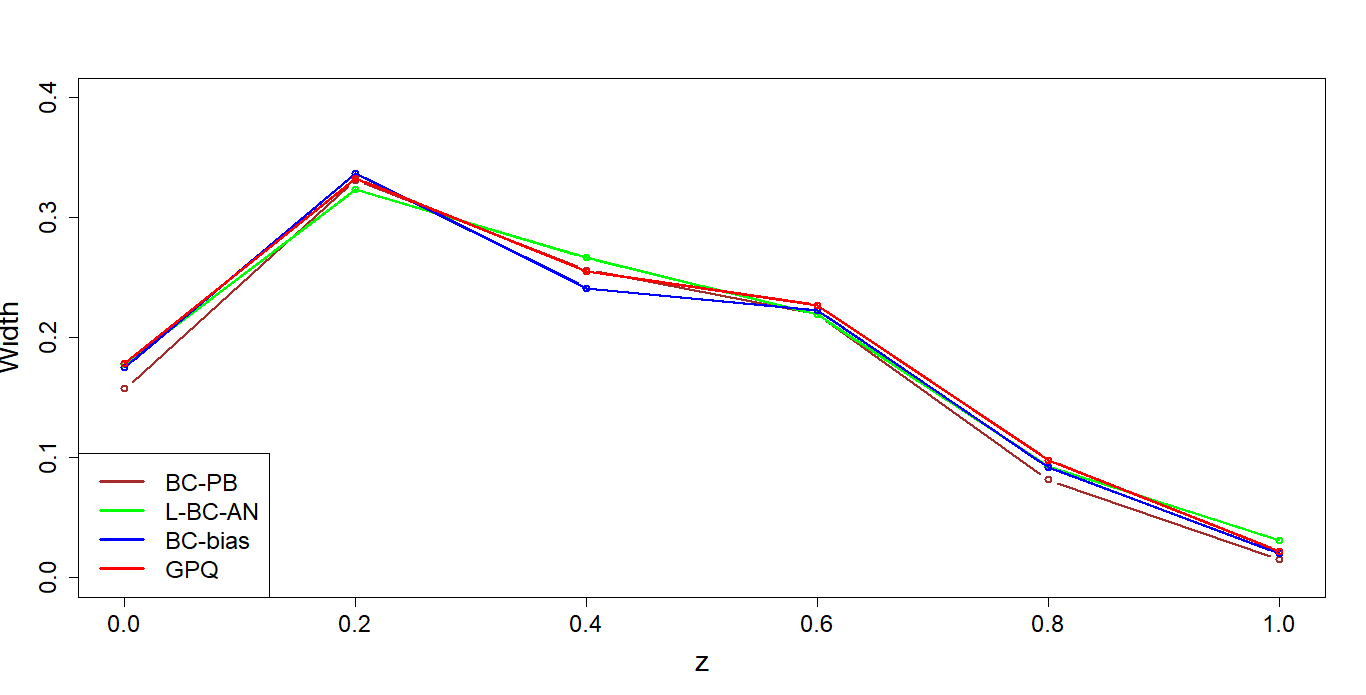}
      
        \caption*{(50,50)}
    \end{subfigure}
    \begin{subfigure}{0.49\textwidth}
    \includegraphics[width=\linewidth, height=5cm, keepaspectratio]{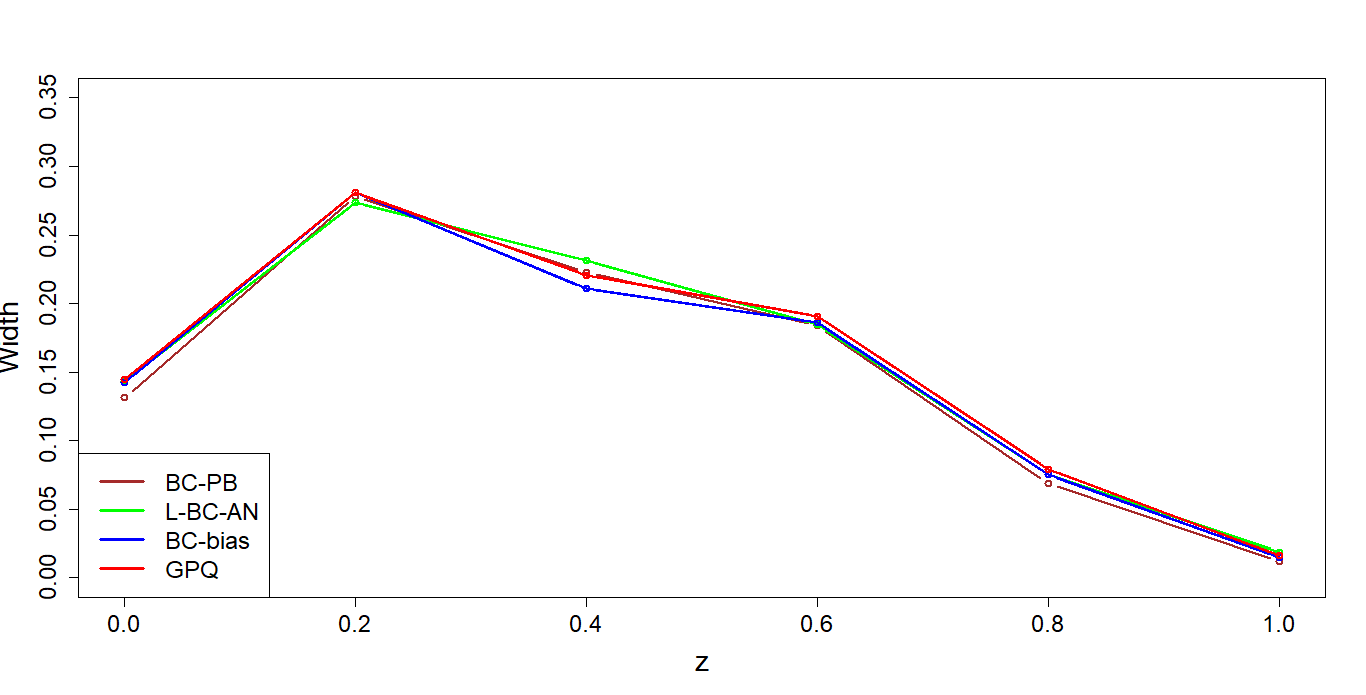}
        \caption*{(50,100)}
    \end{subfigure}
    \begin{subfigure}{0.49\textwidth}
        \includegraphics[width=\linewidth, height=5cm, keepaspectratio]{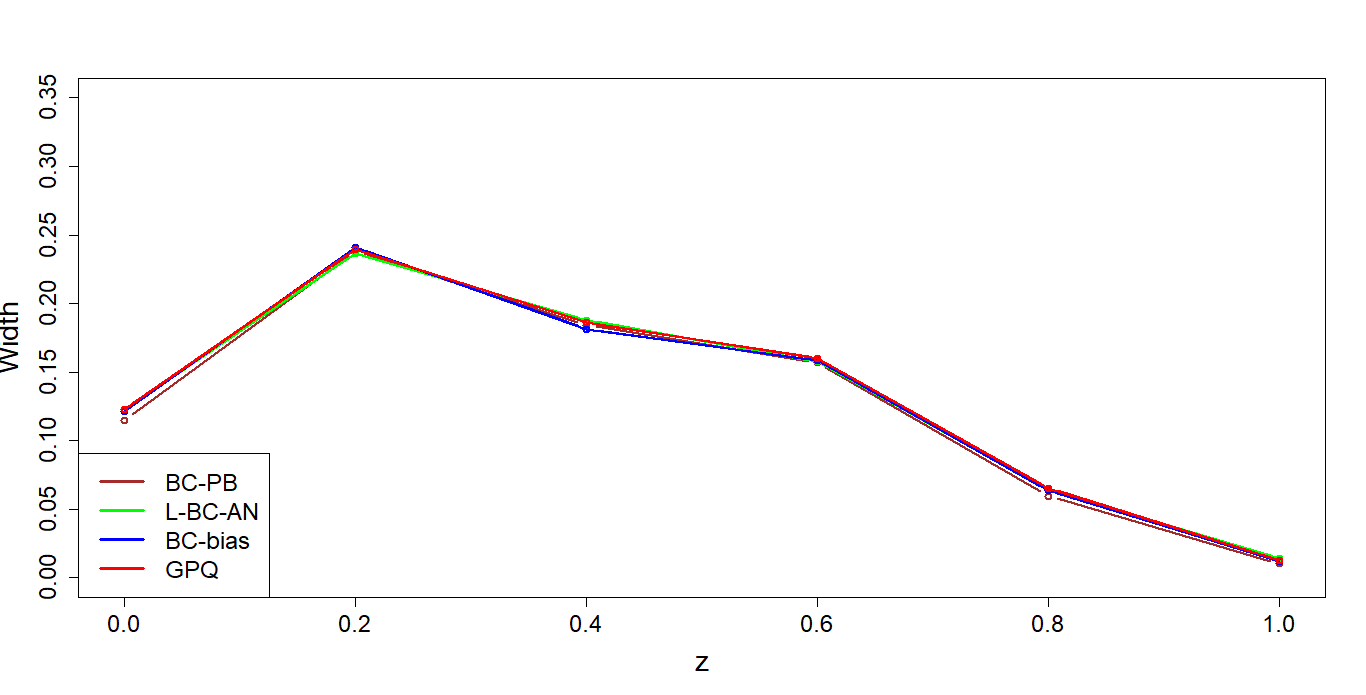}
        \caption*{(100,100)}
    \end{subfigure}
    \begin{subfigure}{0.49\textwidth}
       \includegraphics[width=\linewidth, height=5cm, keepaspectratio]{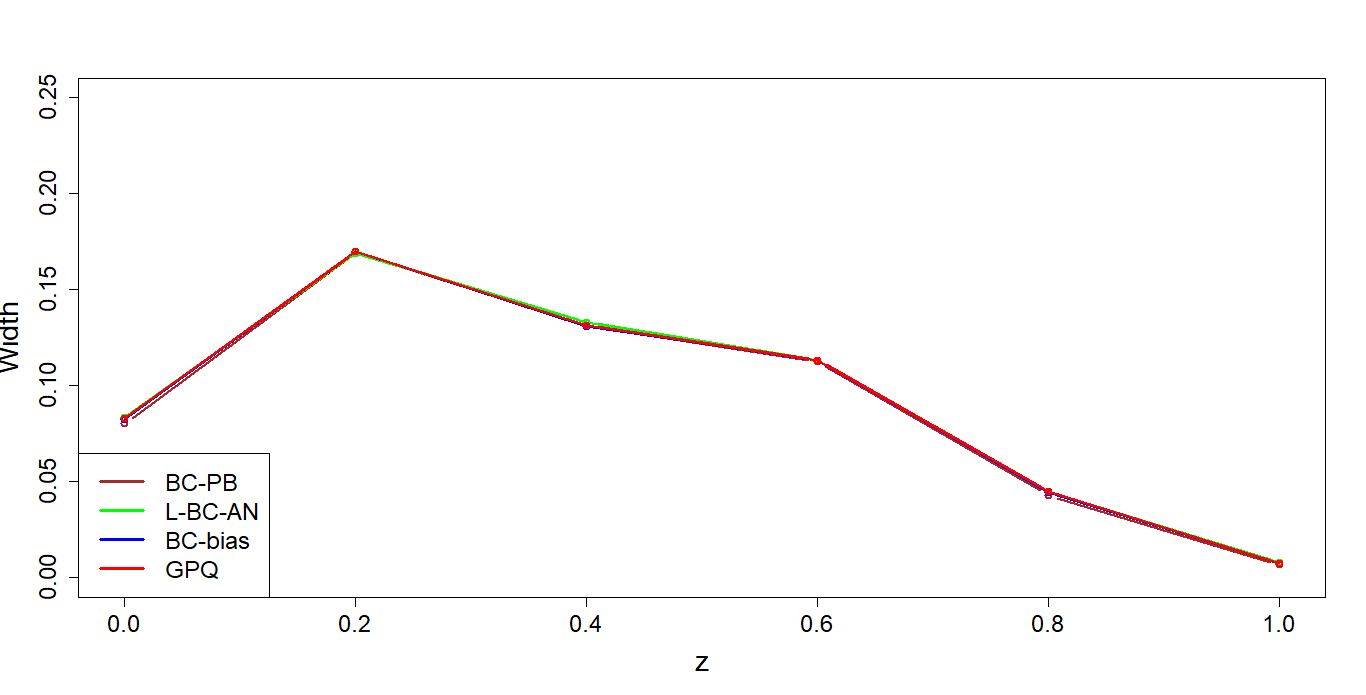}
        \caption*{(200,200)}
    \end{subfigure}
     \caption{Mean interval widths of GPQ and bootstrap methods for the Normal case, scenario 2}
    \label{fig:lennormal2}
\end{figure}

\begin{figure}
    \centering
    \begin{subfigure}{0.49\textwidth}
         \includegraphics[width=\linewidth, height=5cm, keepaspectratio]{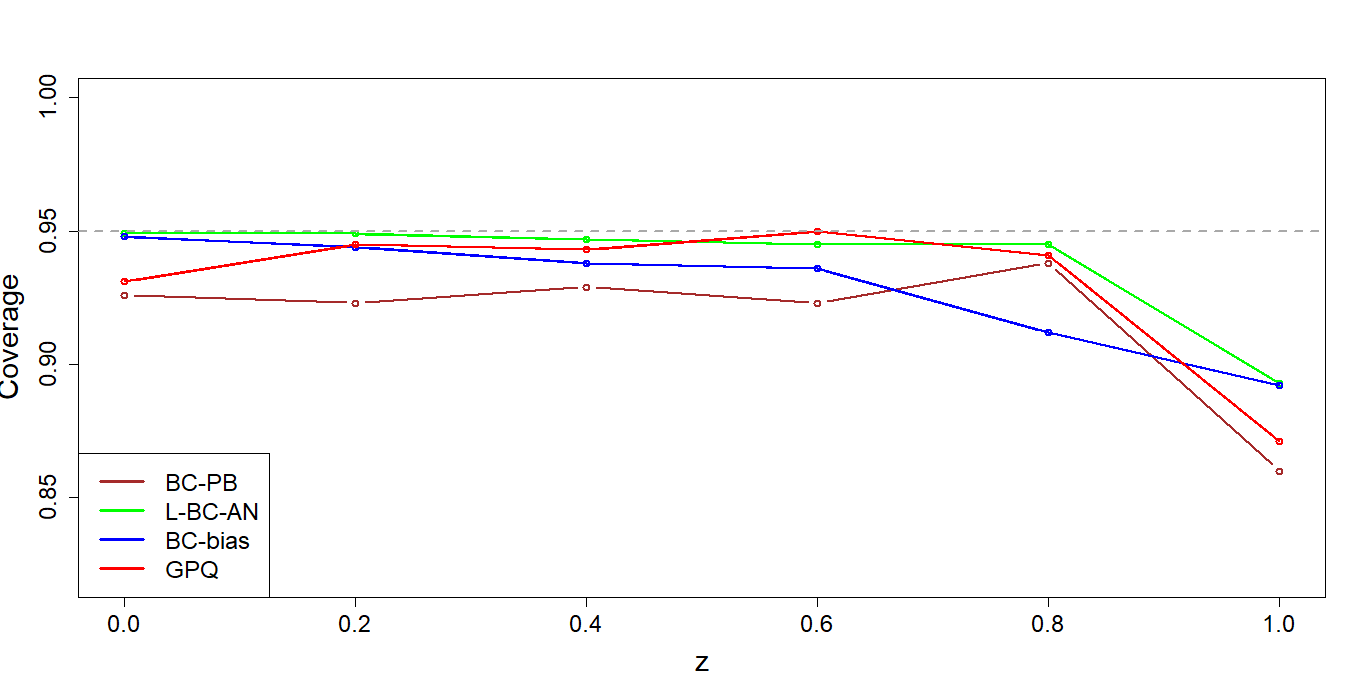}
      
        \caption*{(50,50)}
    \end{subfigure}
    \begin{subfigure}{0.49\textwidth}
    \includegraphics[width=\linewidth, height=5cm, keepaspectratio]{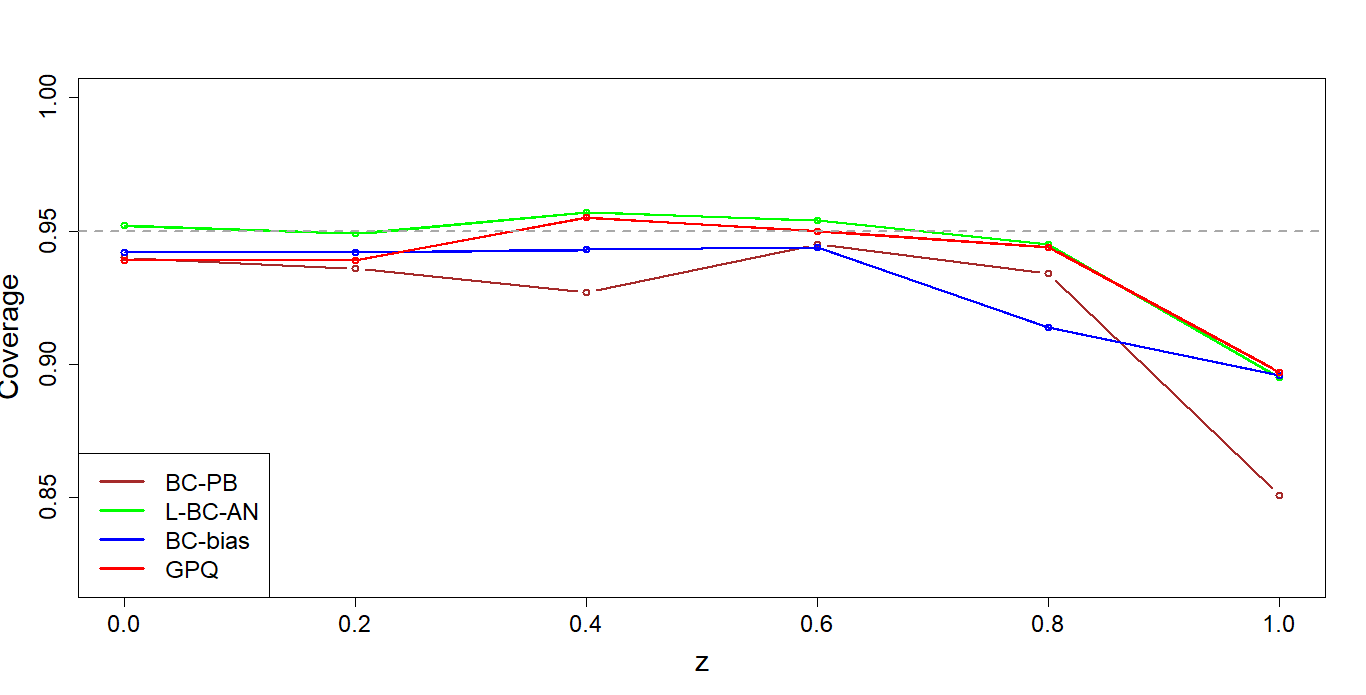}
        \caption*{(50,100)}
    \end{subfigure}
    \begin{subfigure}{0.49\textwidth}
        \includegraphics[width=\linewidth, height=5cm, keepaspectratio]{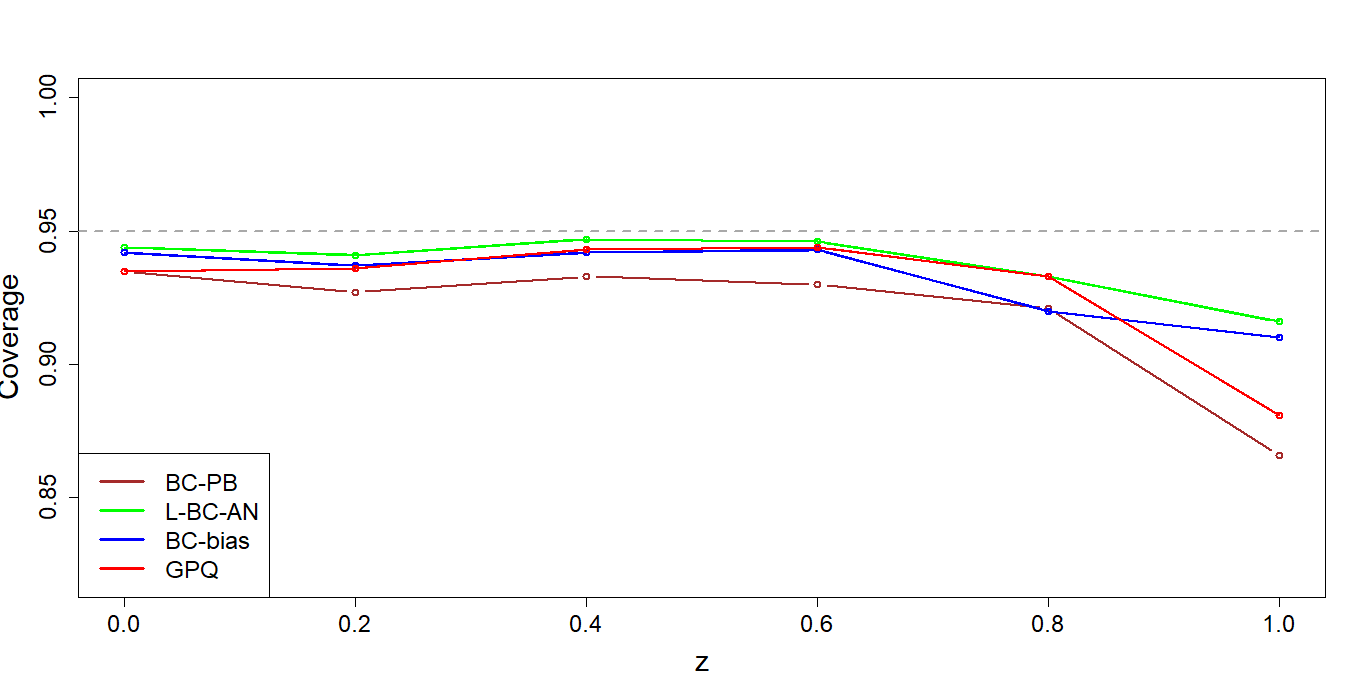}
        \caption*{(100,100)}
    \end{subfigure}
    \begin{subfigure}{0.49\textwidth}
       \includegraphics[width=\linewidth, height=5cm, keepaspectratio]{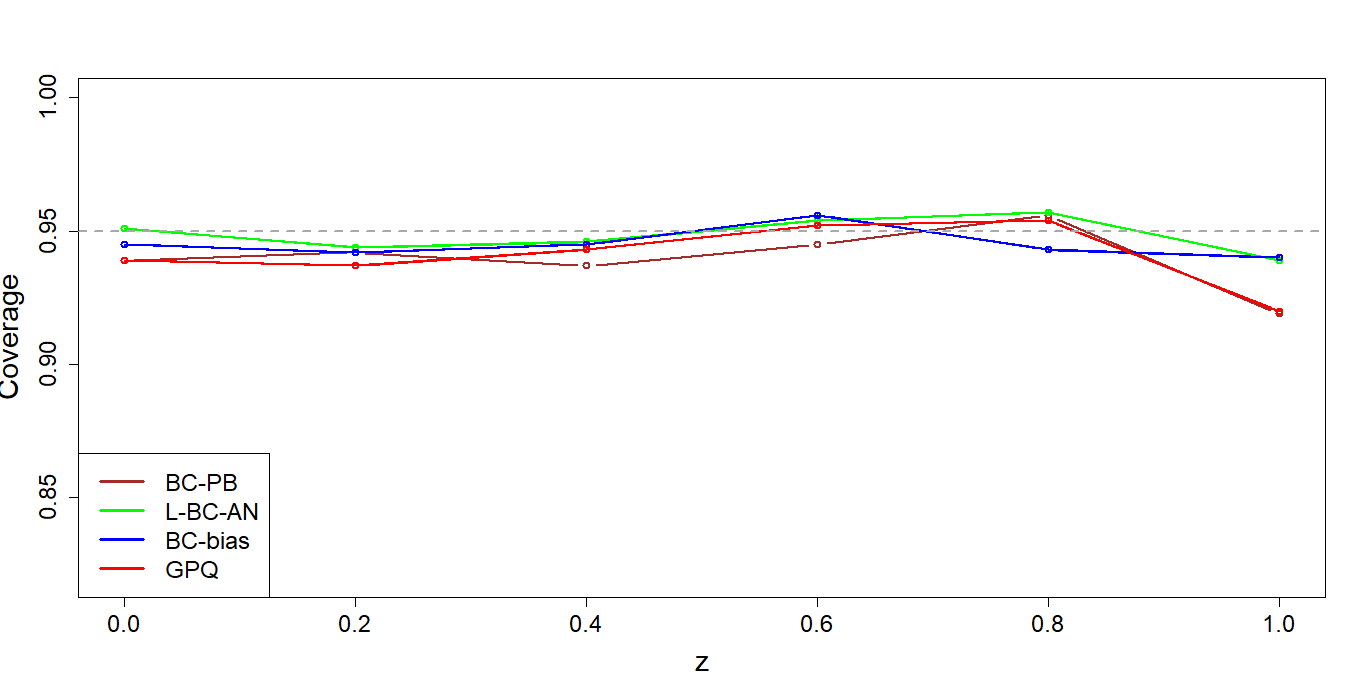}
        \caption*{(200,200)}
    \end{subfigure}
     \caption{Coverages of GPQ and bootstrap methods for the LogNormal case, scenario 1}
    \label{fig:CPLN1}
\end{figure}

\begin{figure}
    \centering
    \begin{subfigure}{0.49\textwidth}
         \includegraphics[width=\linewidth, height=5cm, keepaspectratio]{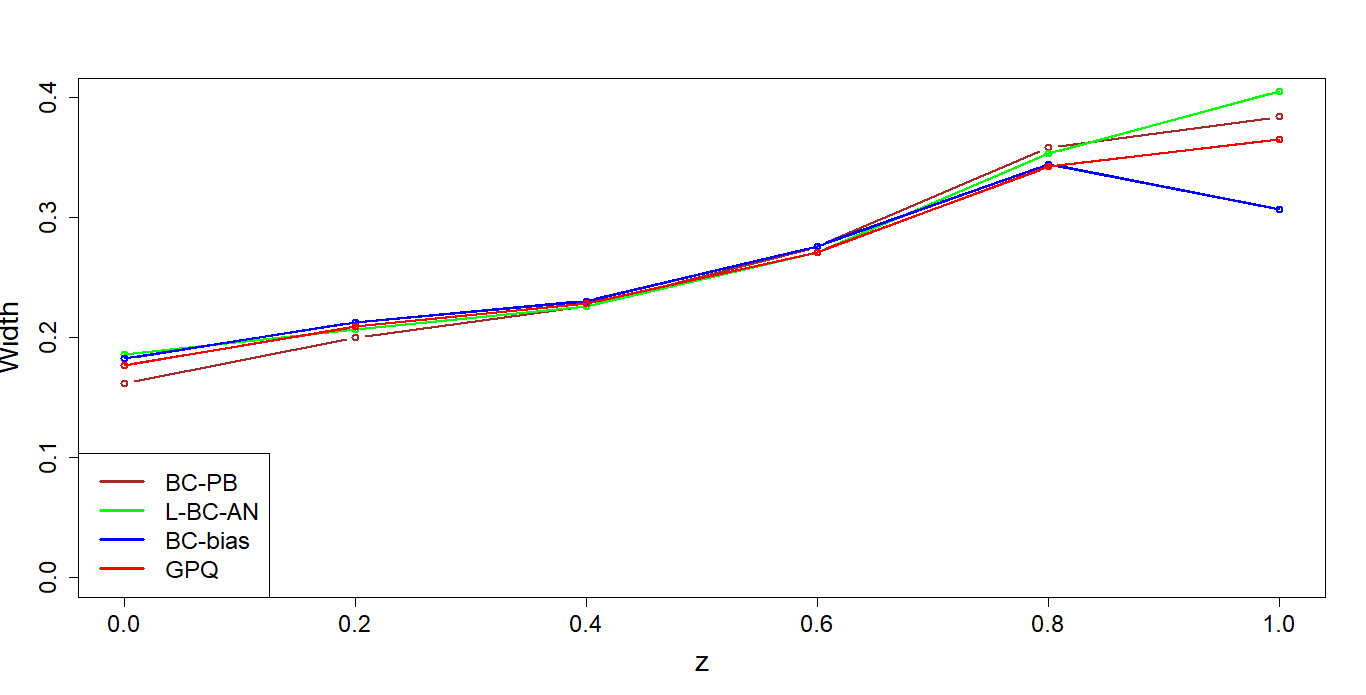}
      
        \caption*{(50,50)}
    \end{subfigure}
    \begin{subfigure}{0.49\textwidth}
    \includegraphics[width=\linewidth, height=5cm, keepaspectratio]{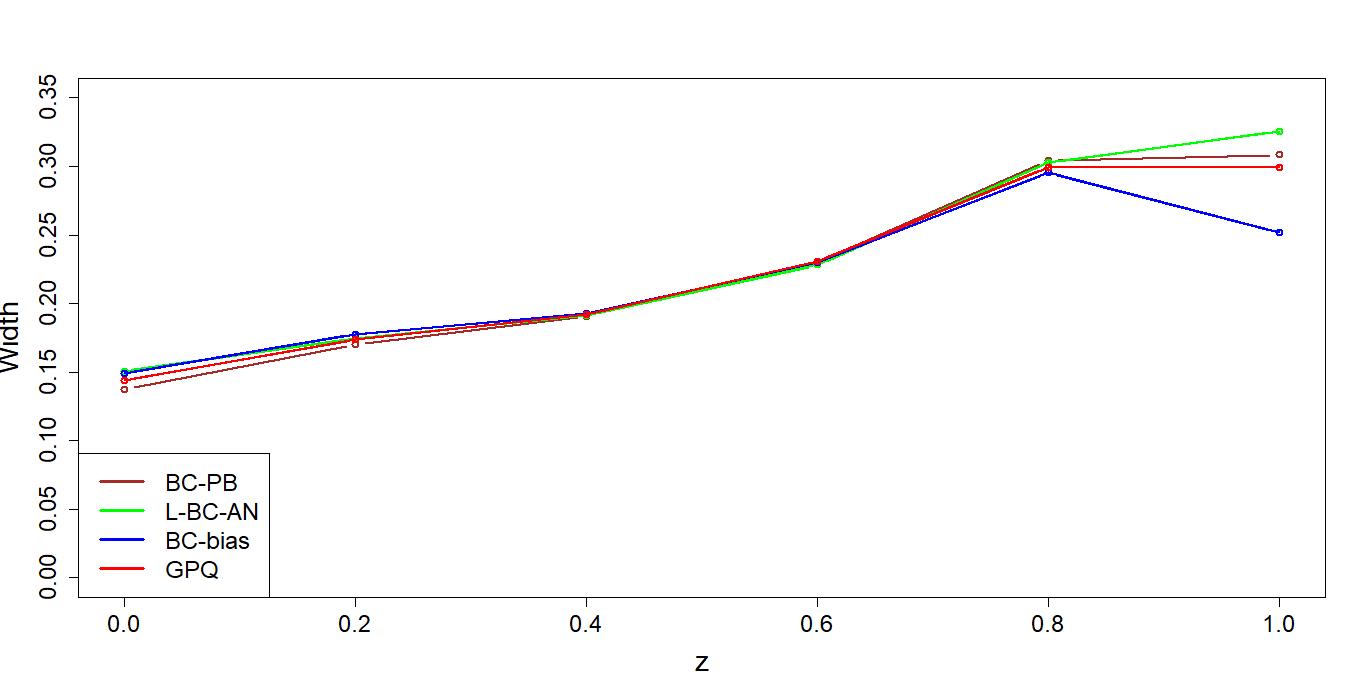}
        \caption*{(50,100)}
    \end{subfigure}
    \begin{subfigure}{0.49\textwidth}
        \includegraphics[width=\linewidth, height=5cm, keepaspectratio]{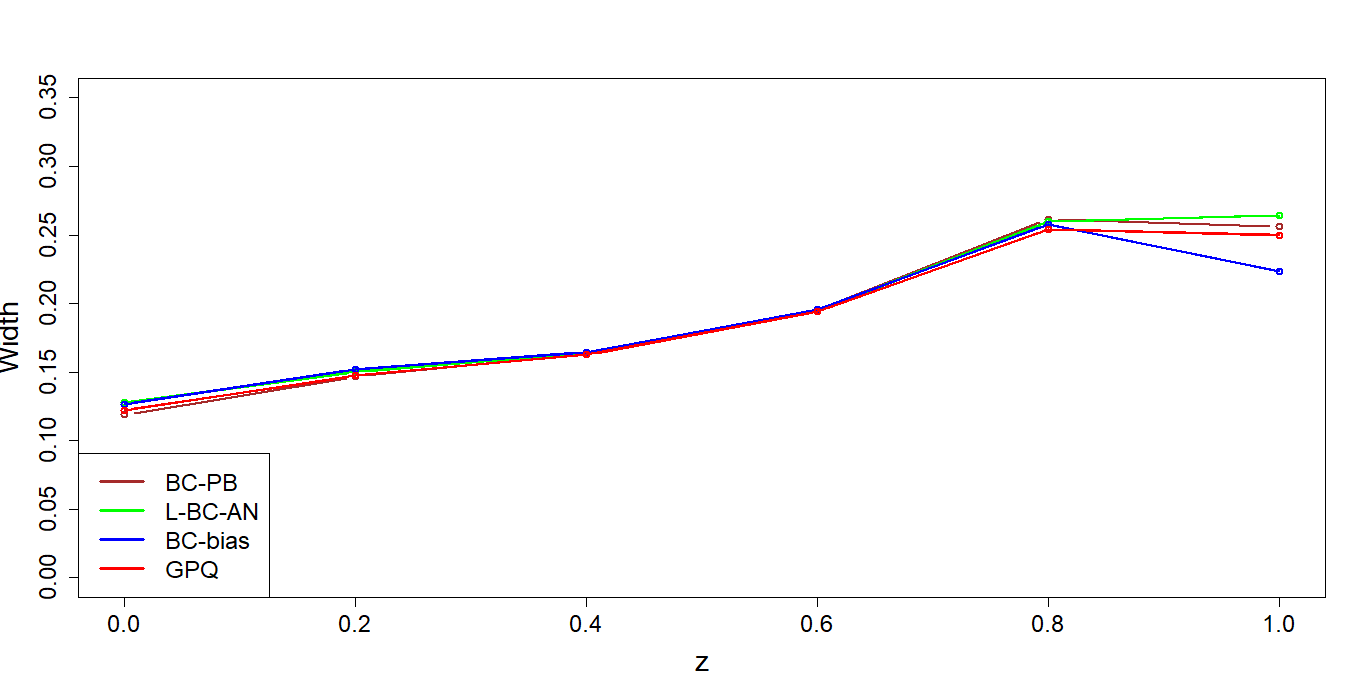}
        \caption*{(100,100)}
    \end{subfigure}
    \begin{subfigure}{0.49\textwidth}
       \includegraphics[width=\linewidth, height=5cm, keepaspectratio]{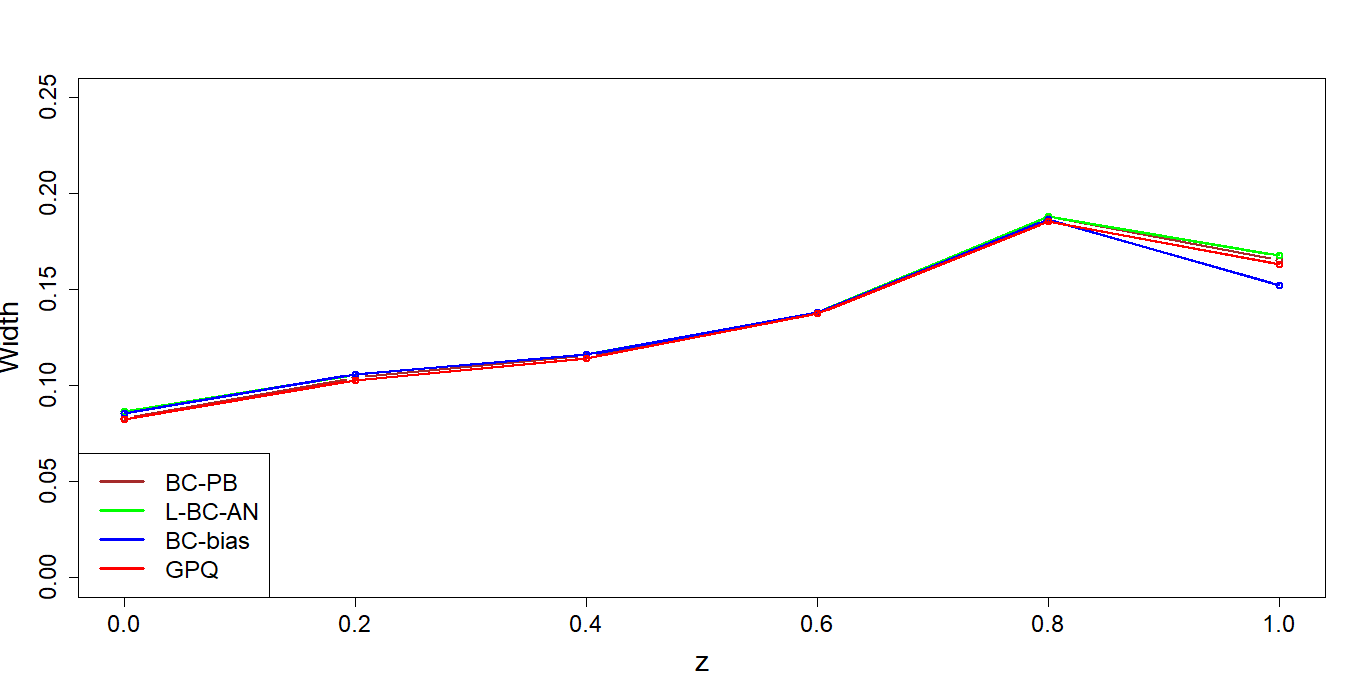}
        \caption*{(200,200)}
    \end{subfigure}
     \caption{Mean interval widths of GPQ and bootstrap methods for the LogNormal case, scenario 1}
    \label{fig:LENLN1}
\end{figure}

\begin{figure}
    \centering
    \begin{subfigure}{0.49\textwidth}
         \includegraphics[width=\linewidth, height=5cm, keepaspectratio]{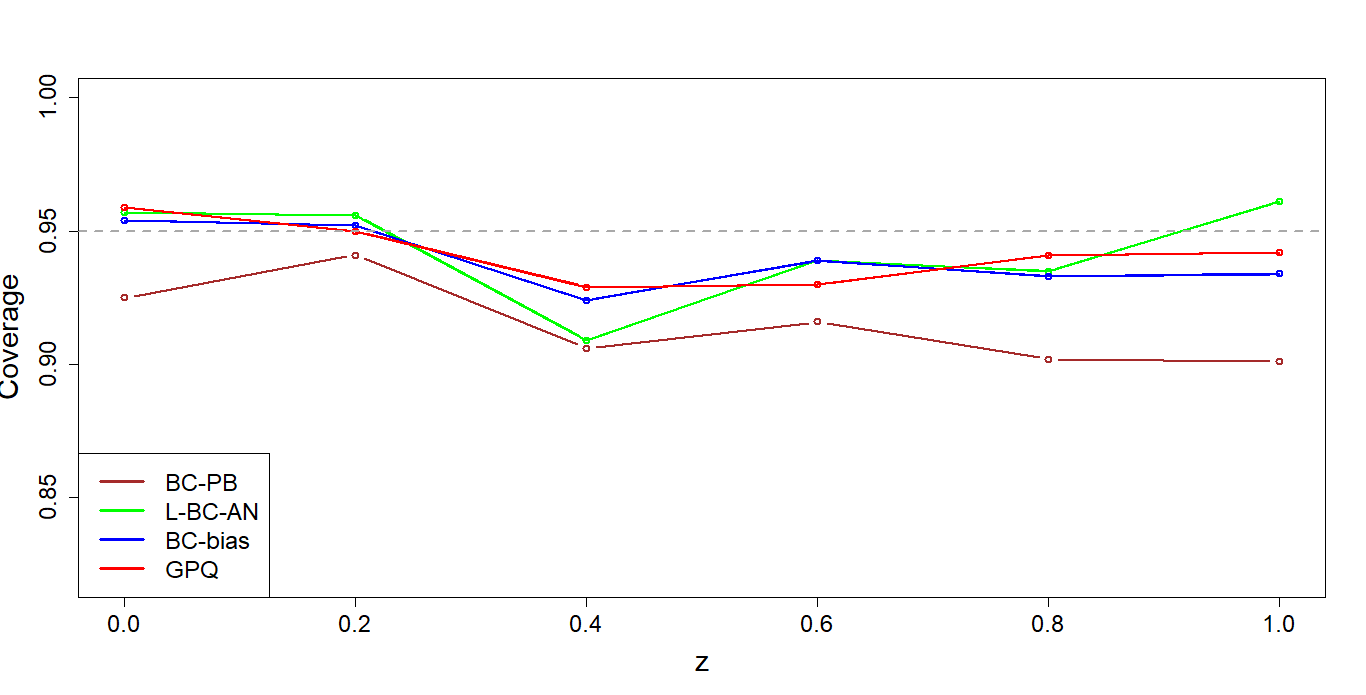}
      
        \caption*{(50,50)}
    \end{subfigure}
    \begin{subfigure}{0.49\textwidth}
    \includegraphics[width=\linewidth, height=5cm, keepaspectratio]{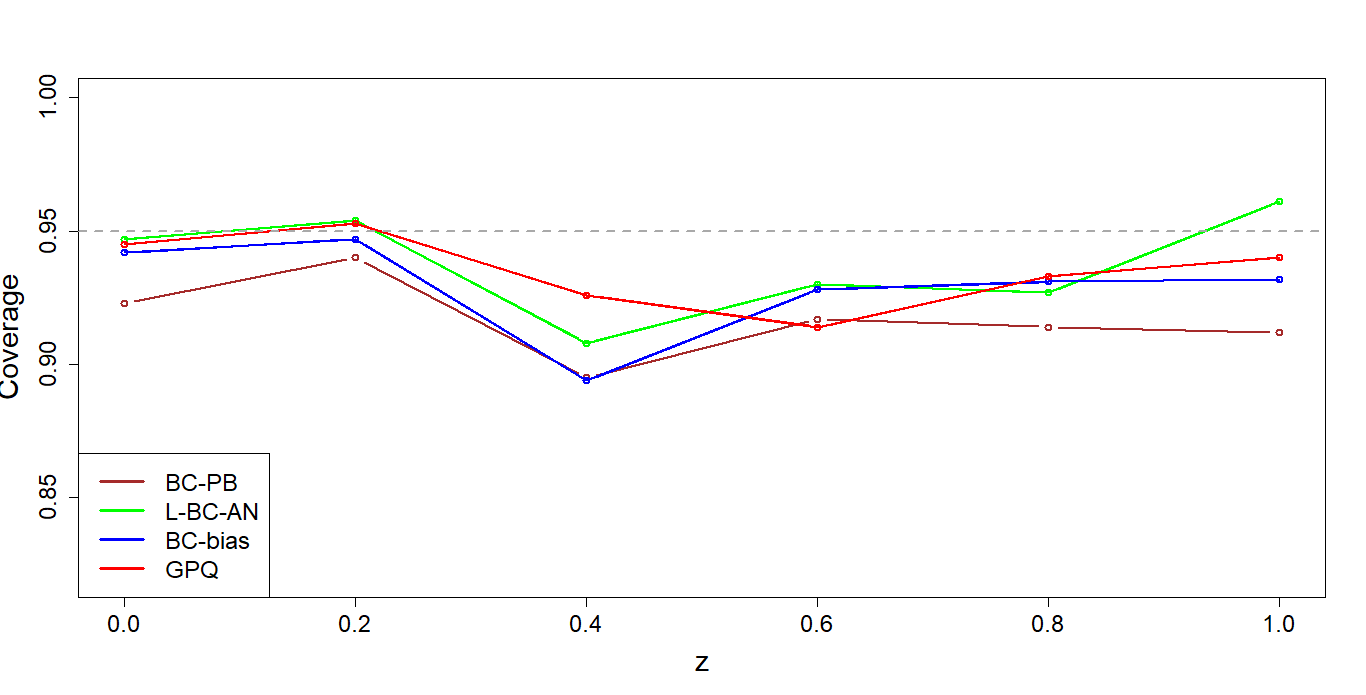}
        \caption*{(50,100)}
    \end{subfigure}
    \begin{subfigure}{0.49\textwidth}
        \includegraphics[width=\linewidth, height=5cm, keepaspectratio]{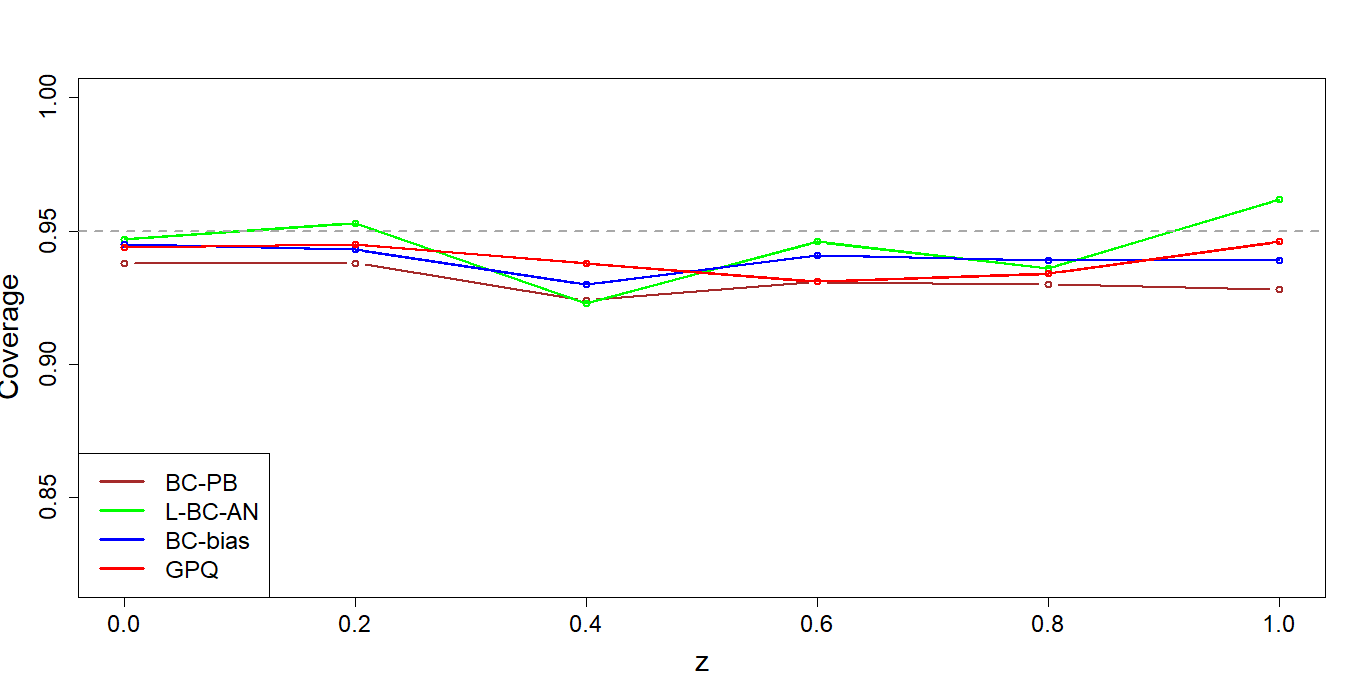}
        \caption*{(100,100)}
    \end{subfigure}
    \begin{subfigure}{0.49\textwidth}
       \includegraphics[width=\linewidth, height=5cm, keepaspectratio]{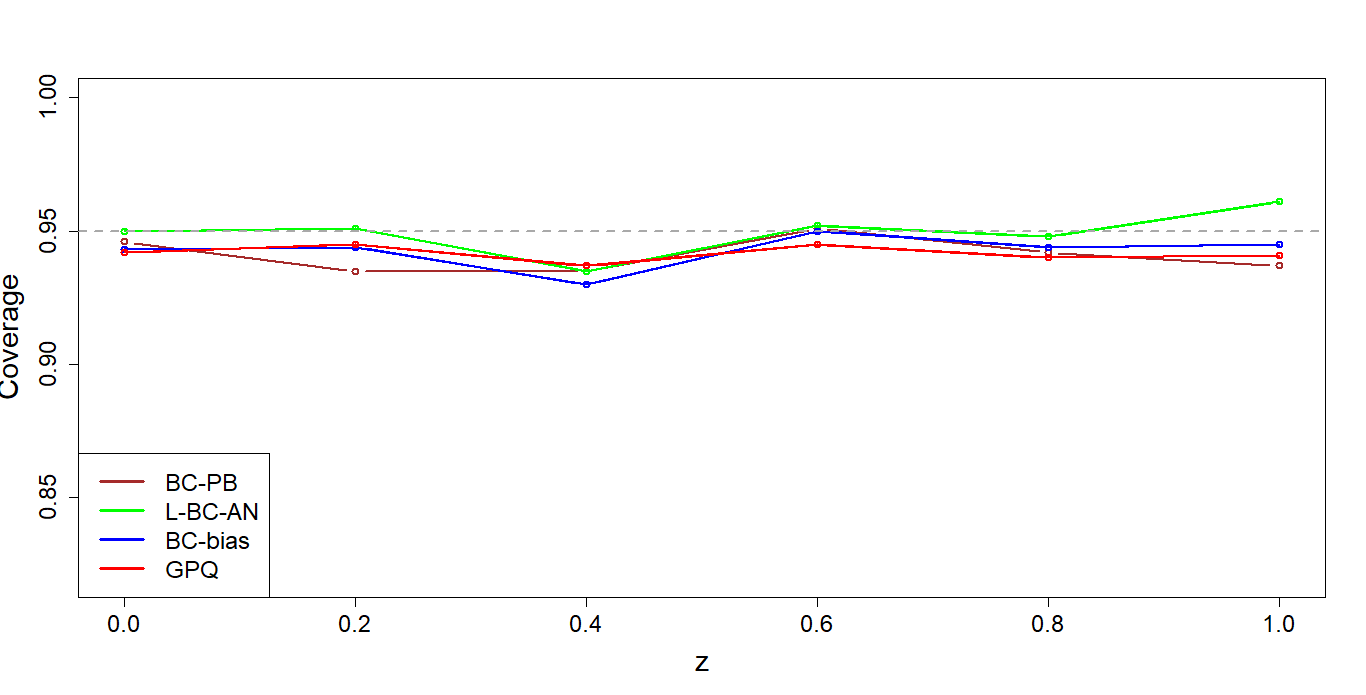}
        \caption*{(200,200)}
    \end{subfigure}
     \caption{Coverages of GPQ and bootstrap methods for the LogNormal case, scenario 2}
    \label{fig:CPLN2}
\end{figure}

\begin{figure}
    \centering
    \begin{subfigure}{0.49\textwidth}
         \includegraphics[width=\linewidth, height=5cm, keepaspectratio]{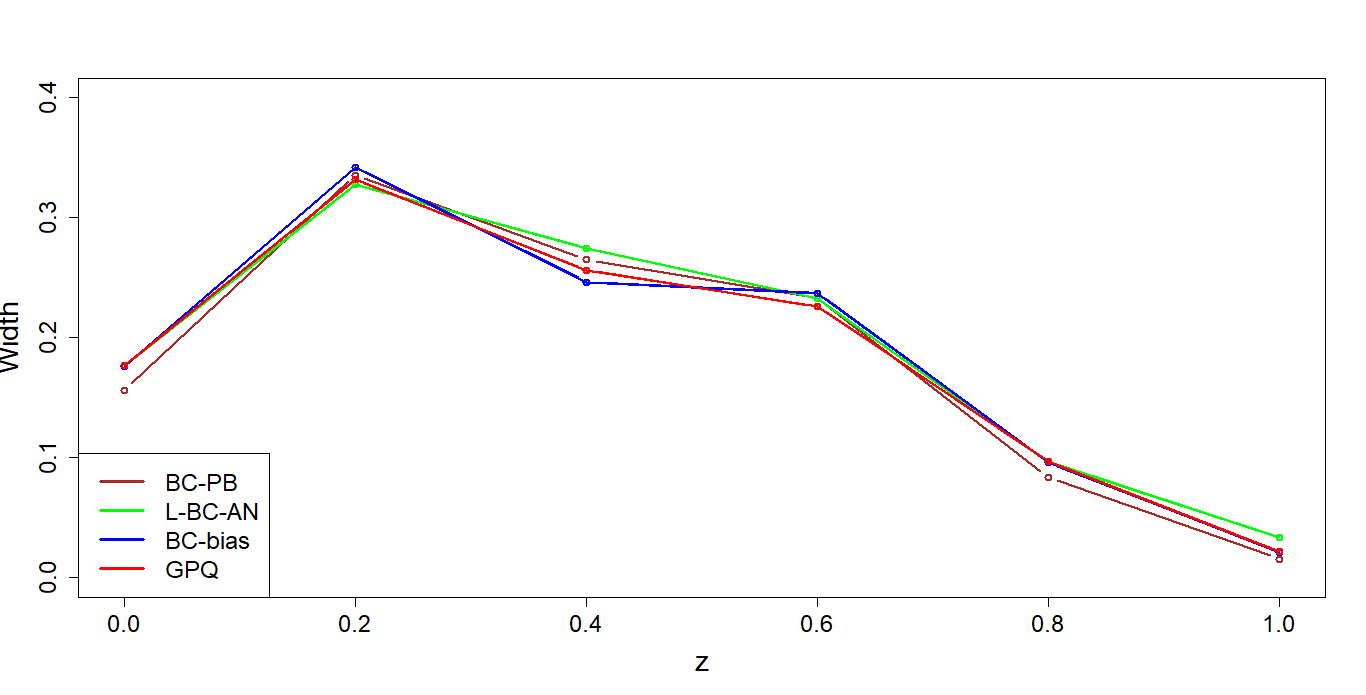}
      
        \caption*{(50,50)}
    \end{subfigure}
    \begin{subfigure}{0.49\textwidth}
    \includegraphics[width=\linewidth, height=5cm, keepaspectratio]{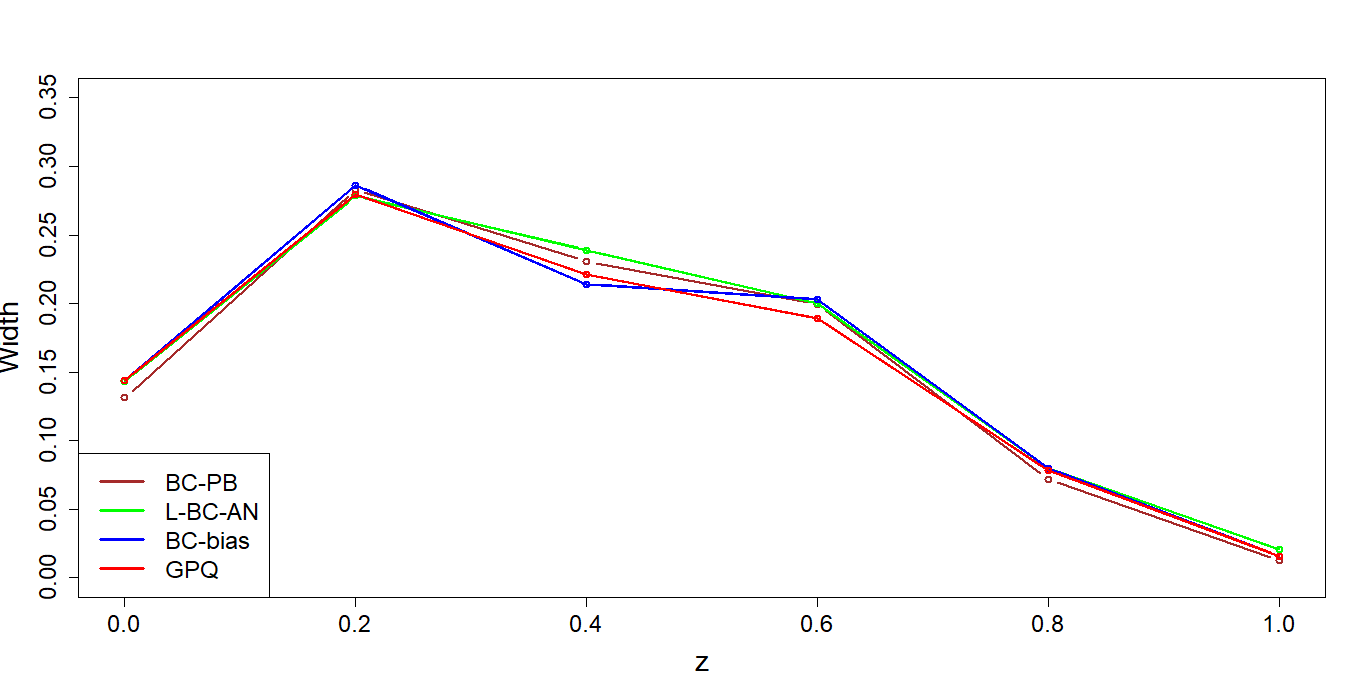}
        \caption*{(50,100)}
    \end{subfigure}
    \begin{subfigure}{0.49\textwidth}
        \includegraphics[width=\linewidth, height=5cm, keepaspectratio]{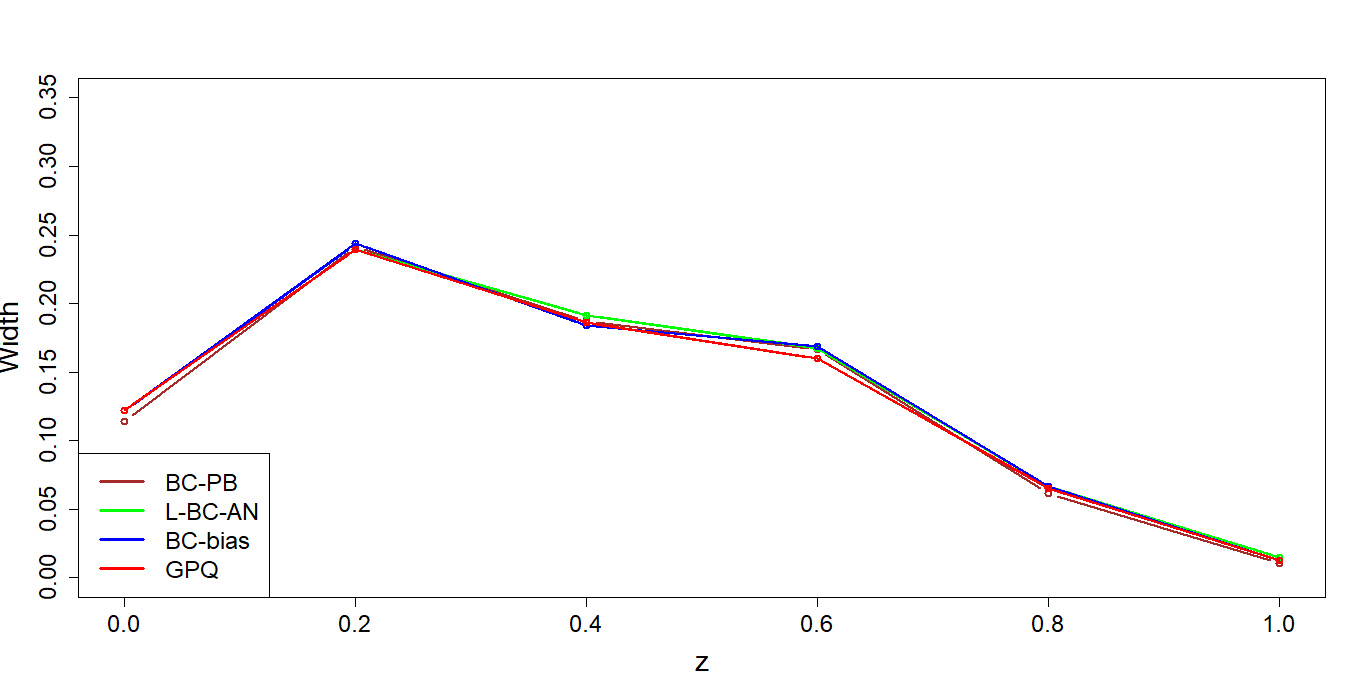}
        \caption*{(100,100)}
    \end{subfigure}
    \begin{subfigure}{0.49\textwidth}
       \includegraphics[width=\linewidth, height=5cm, keepaspectratio]{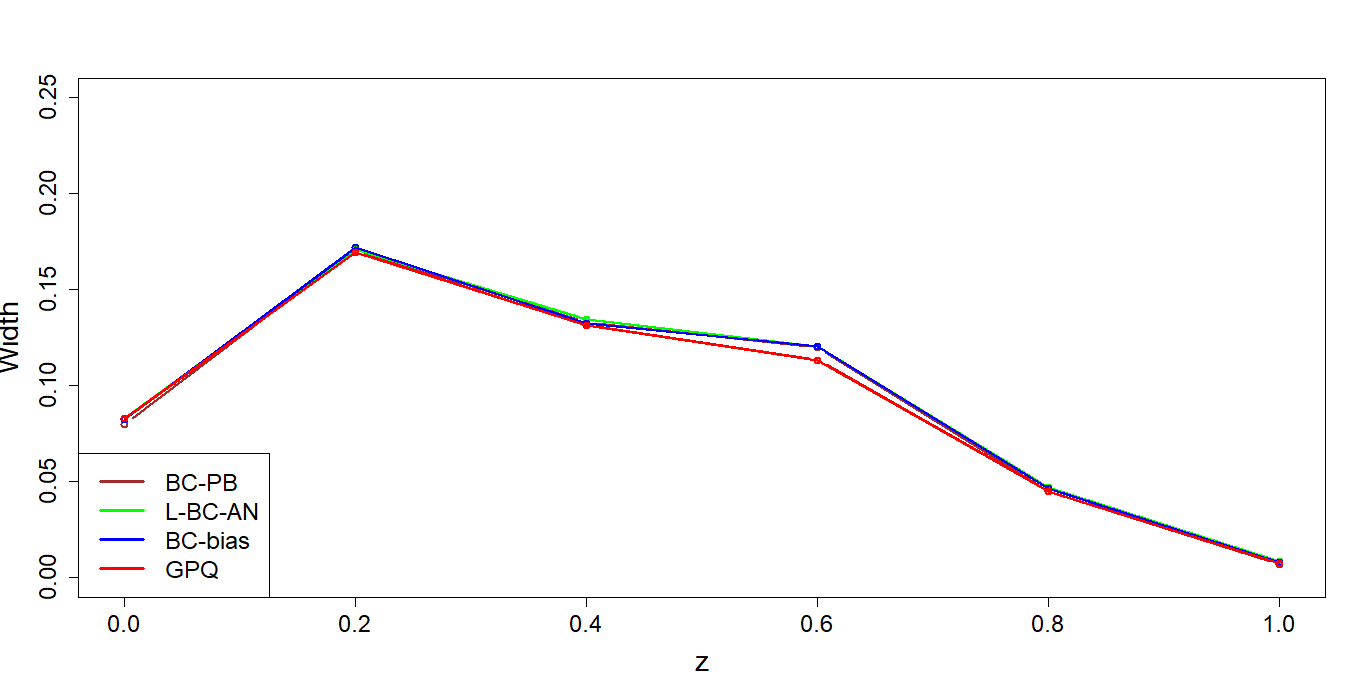}
        \caption*{(200,200)}
    \end{subfigure}
     \caption{Mean interval widths of GPQ and bootstrap methods for the LogNormal case, scenario 2}
    \label{fig:LENLN2}
\end{figure}

\begin{figure}
    \centering
    \begin{subfigure}{0.49\textwidth}
         \includegraphics[width=\linewidth, height=5cm, keepaspectratio]{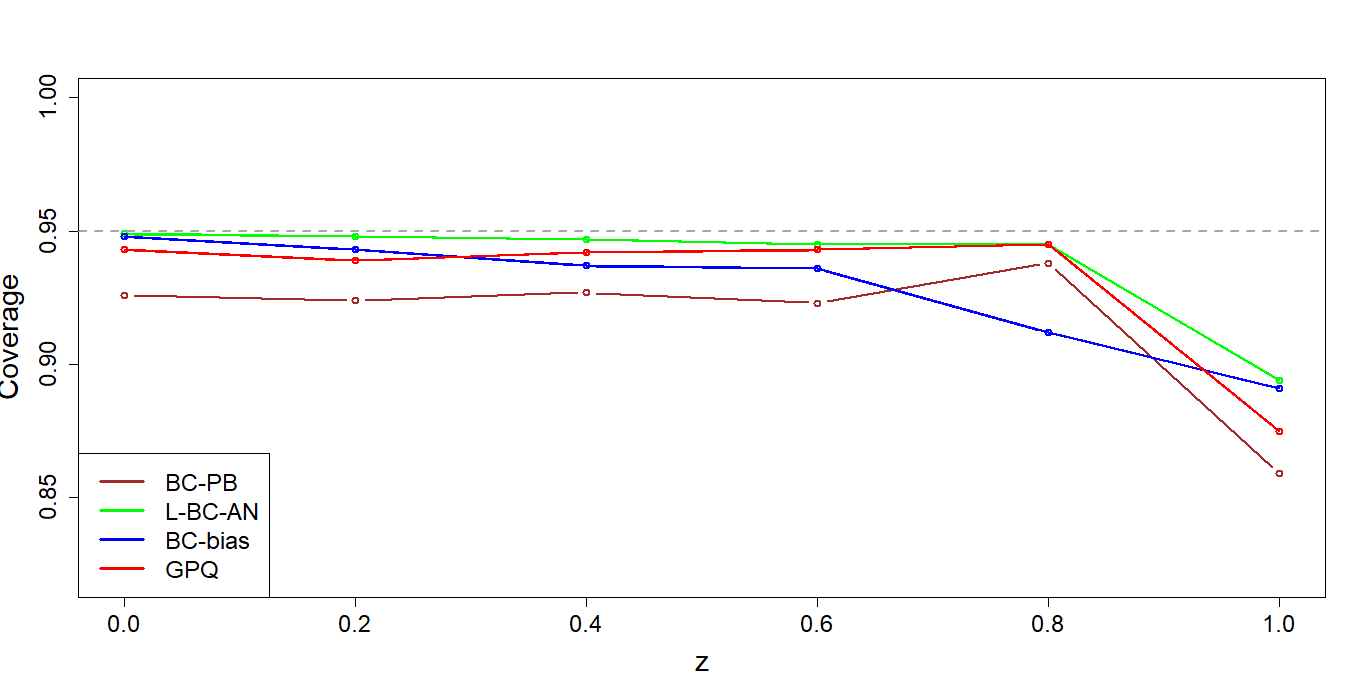}
      
        \caption*{(50,50)}
    \end{subfigure}
    \begin{subfigure}{0.49\textwidth}
    \includegraphics[width=\linewidth, height=5cm, keepaspectratio]{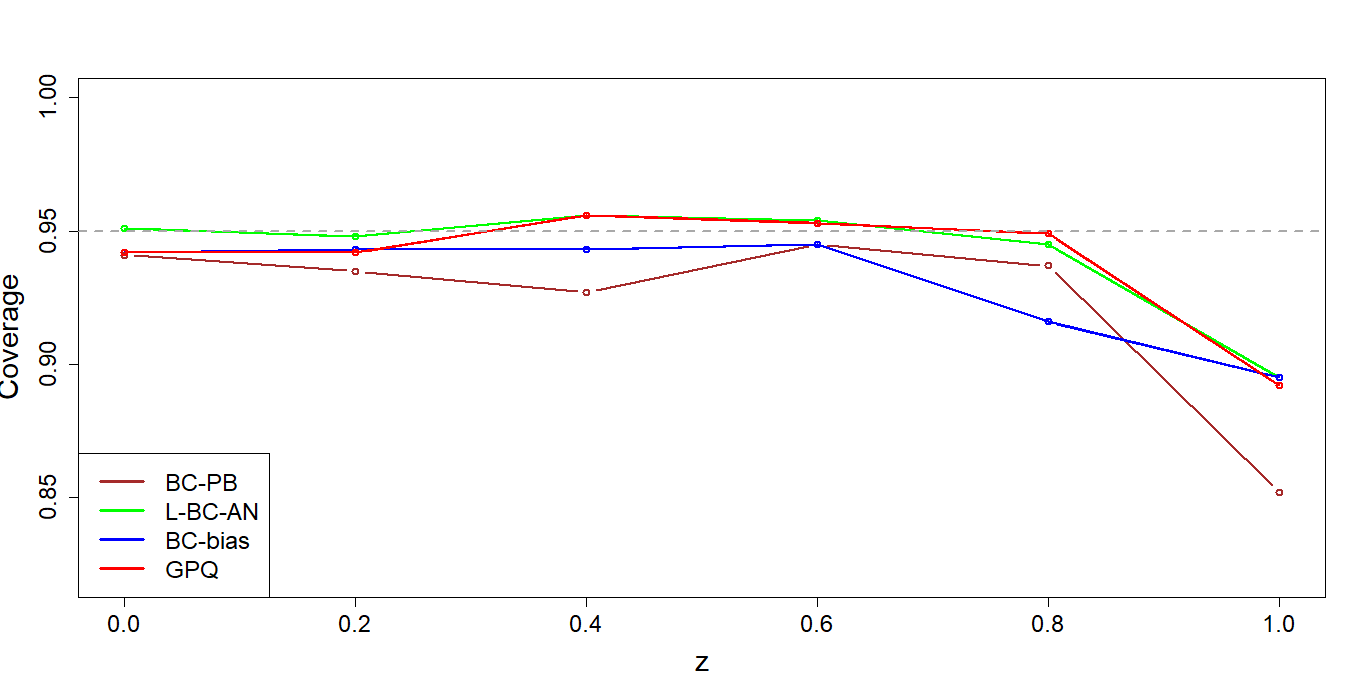}
        \caption*{(50,100)}
    \end{subfigure}
    \begin{subfigure}{0.49\textwidth}
        \includegraphics[width=\linewidth, height=5cm, keepaspectratio]{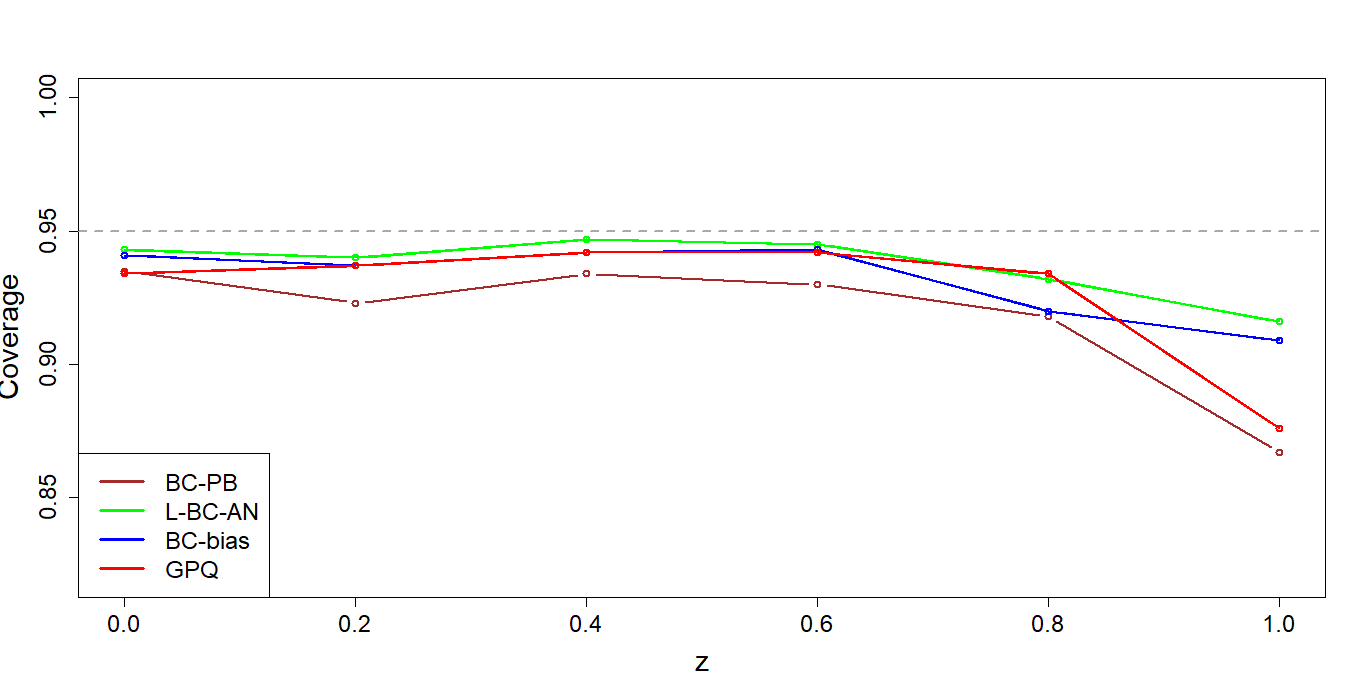}
        \caption*{(100,100)}
    \end{subfigure}
    \begin{subfigure}{0.49\textwidth}
       \includegraphics[width=\linewidth, height=5cm, keepaspectratio]{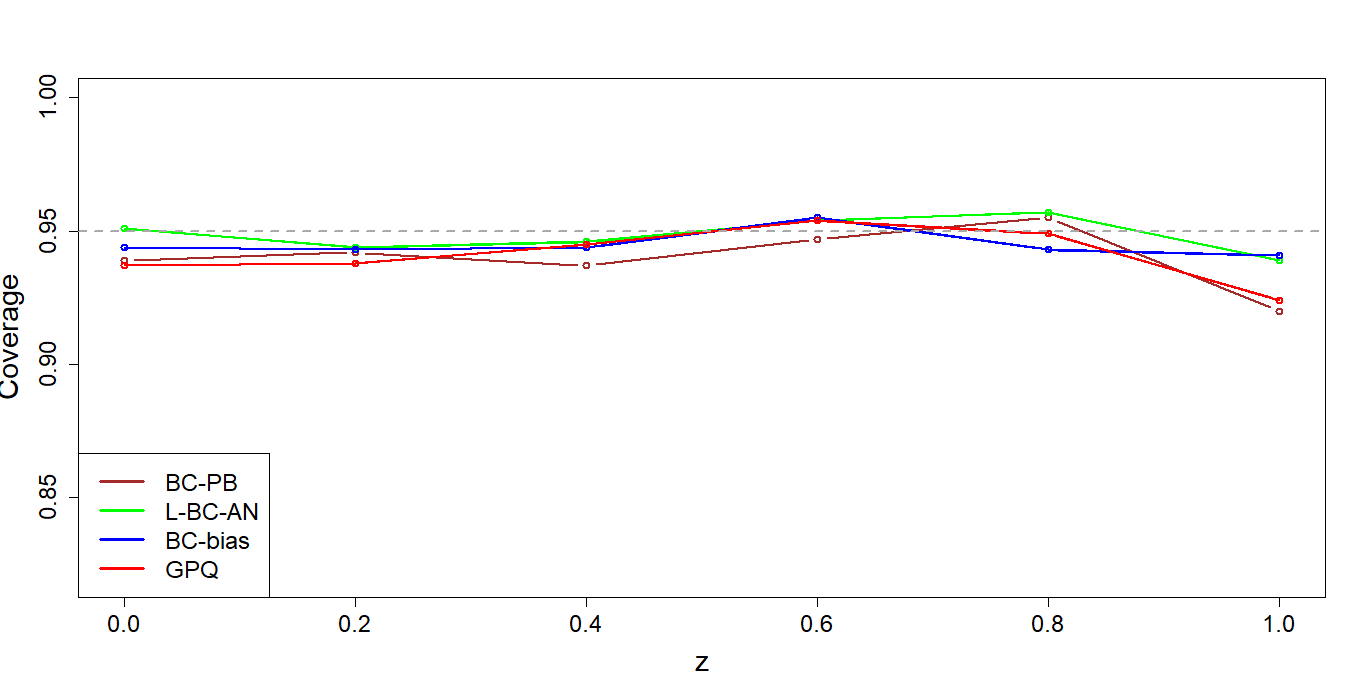}
        \caption*{(200,200)}
    \end{subfigure}
     \caption{Coverages of GPQ and bootstrap methods for the PowerNormal case, scenario 1}
    \label{fig:CPPW1}
\end{figure}

\begin{figure}
    \centering
    \begin{subfigure}{0.49\textwidth}
         \includegraphics[width=\linewidth, height=5cm, keepaspectratio]{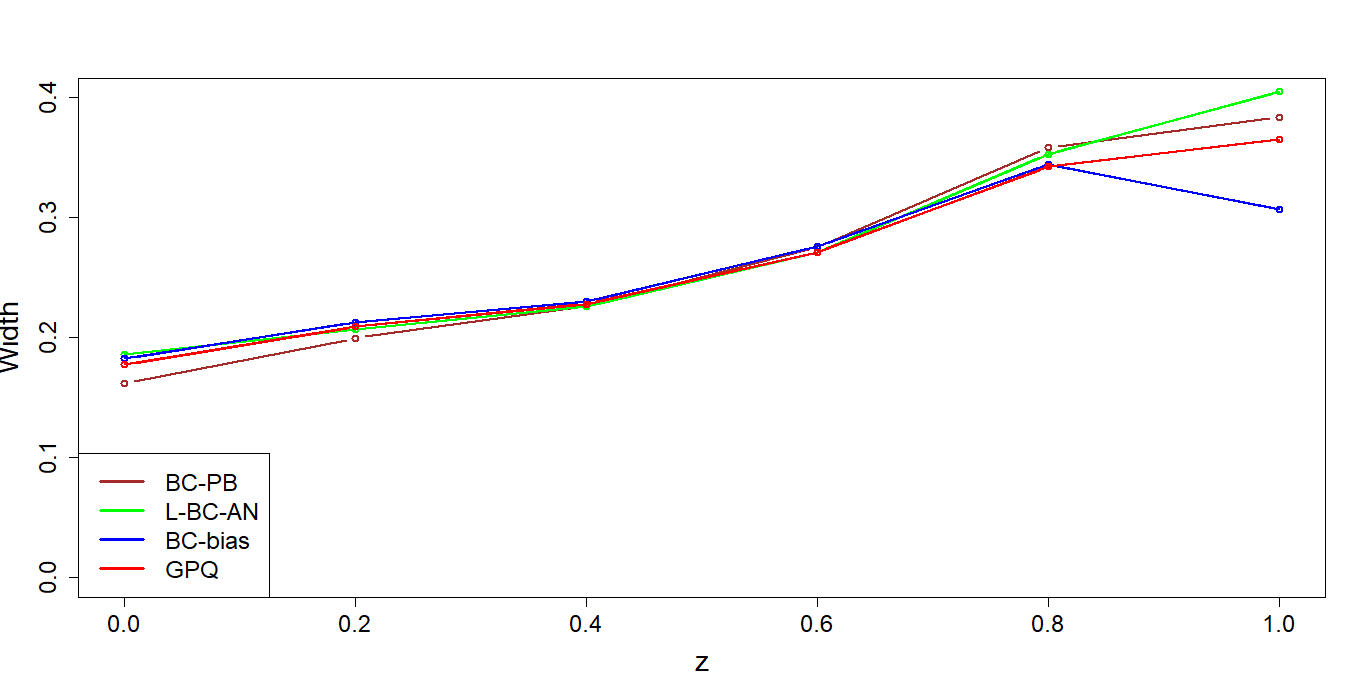}
      
        \caption*{(50,50)}
    \end{subfigure}
    \begin{subfigure}{0.49\textwidth}
    \includegraphics[width=\linewidth, height=5cm, keepaspectratio]{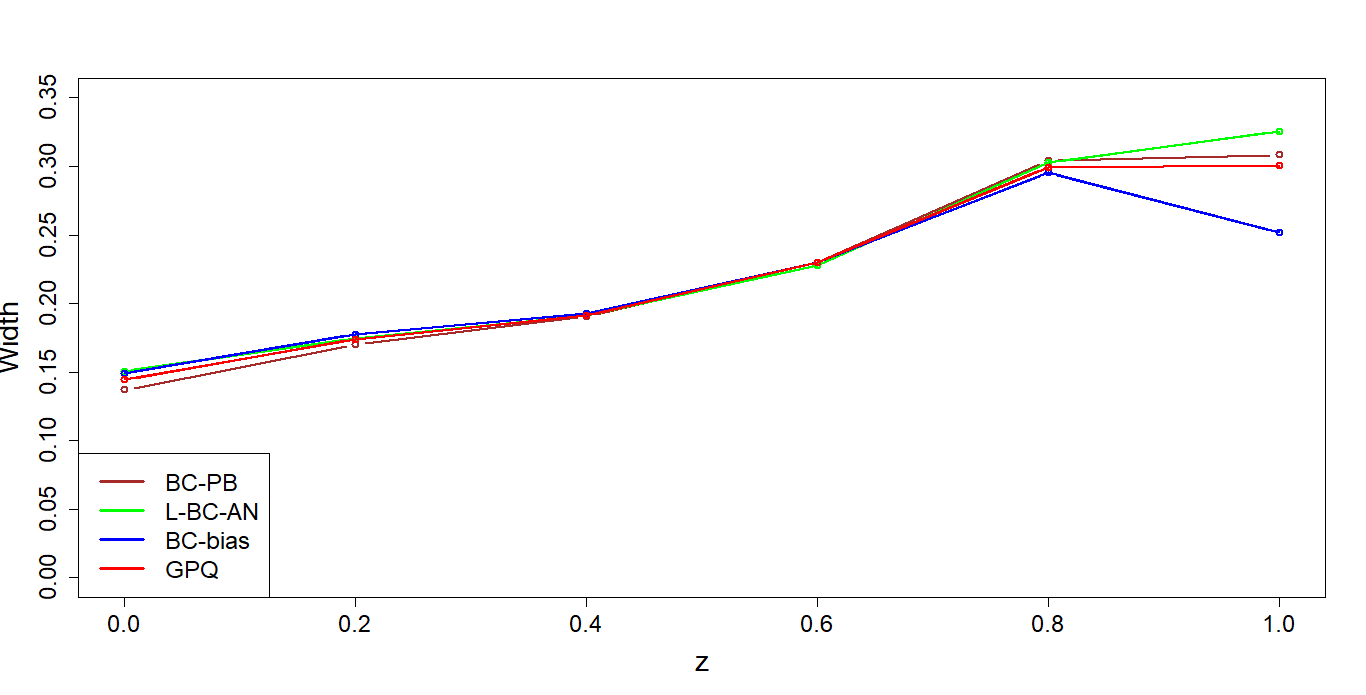}
        \caption*{(50,100)}
    \end{subfigure}
    \begin{subfigure}{0.49\textwidth}
        \includegraphics[width=\linewidth, height=5cm, keepaspectratio]{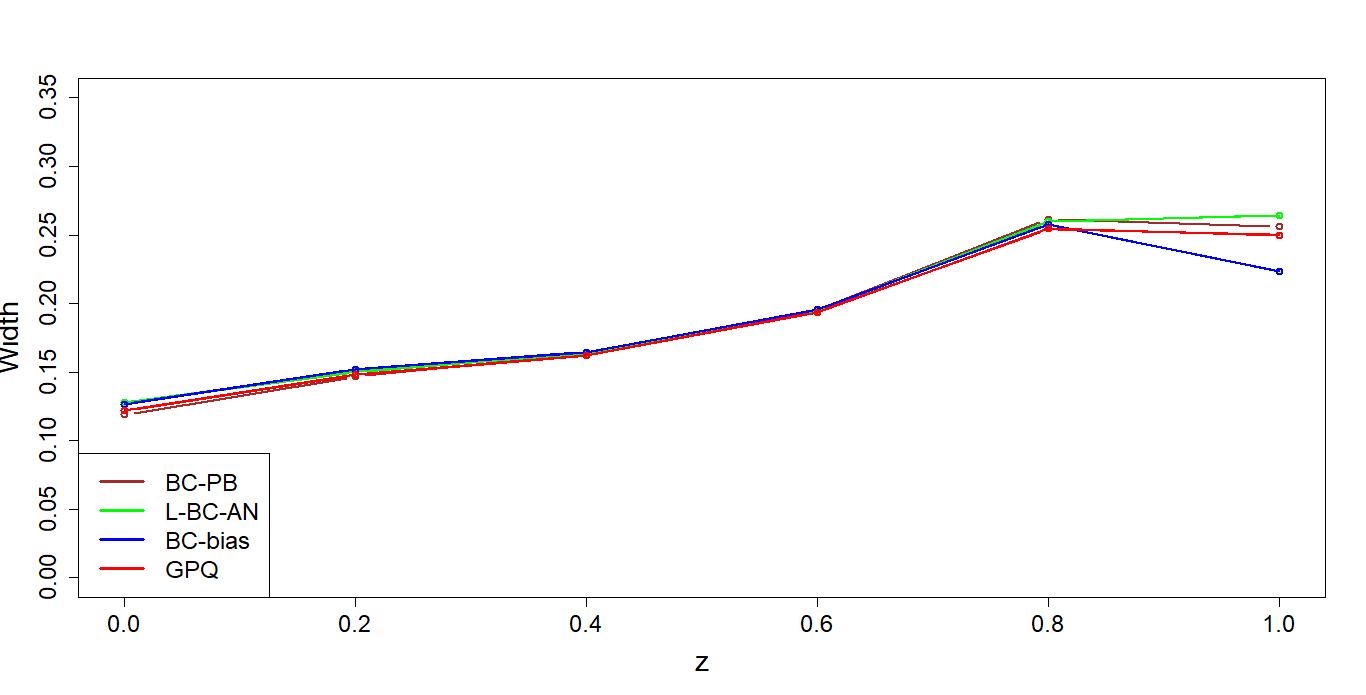}
        \caption*{(100,100)}
    \end{subfigure}
    \begin{subfigure}{0.49\textwidth}
       \includegraphics[width=\linewidth, height=5cm, keepaspectratio]{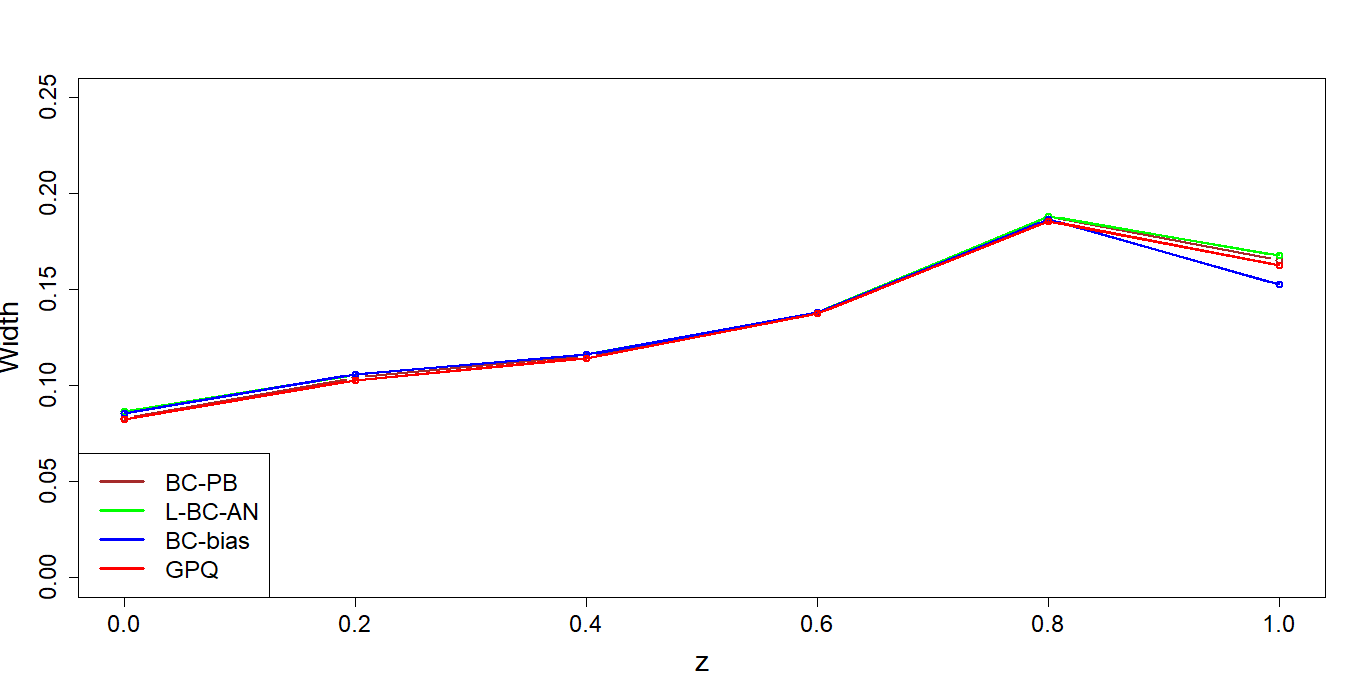}
        \caption*{(200,200)}
    \end{subfigure}
     \caption{Mean interval widths of GPQ and bootstrap methods for the PowerNormal case, scenario 1}
    \label{fig:LENPW1}
\end{figure}

\begin{figure}
    \centering
    \begin{subfigure}{0.49\textwidth}
         \includegraphics[width=\linewidth, height=5cm, keepaspectratio]{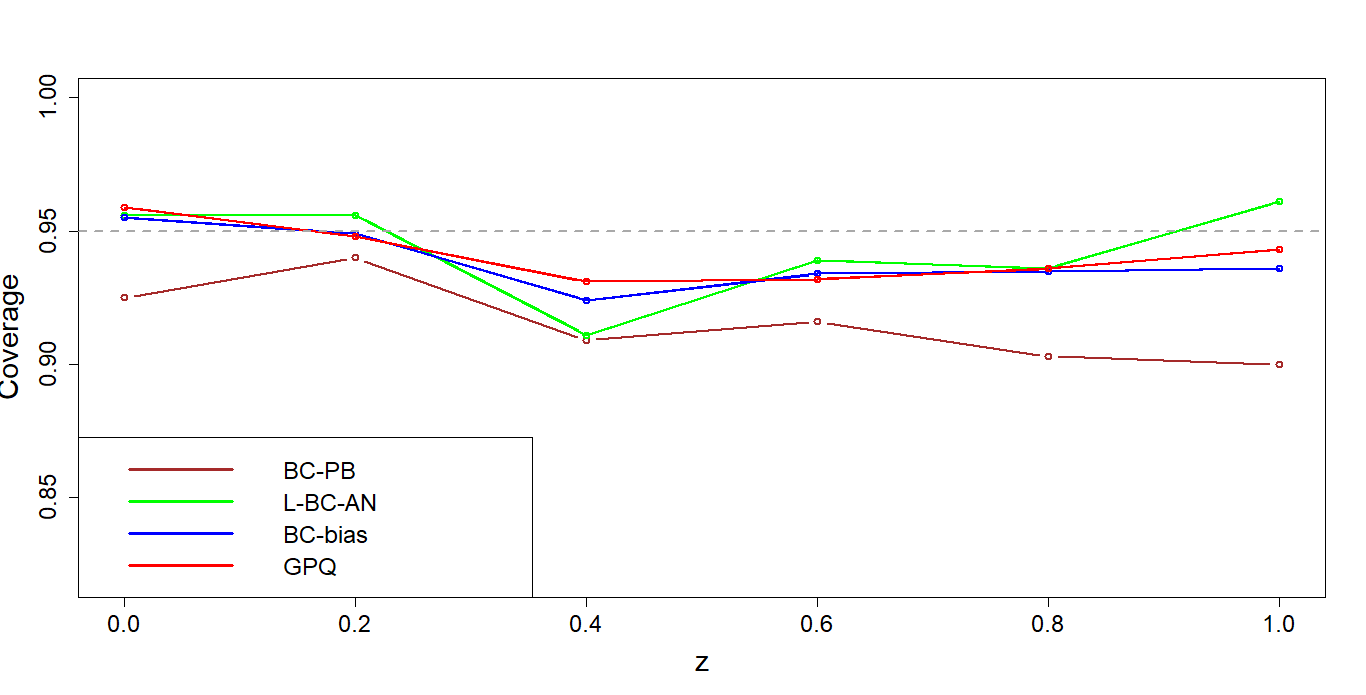}
        \caption*{(50,50)}
    \end{subfigure}
    \begin{subfigure}{0.49\textwidth}
    \includegraphics[width=\linewidth, height=5cm, keepaspectratio]{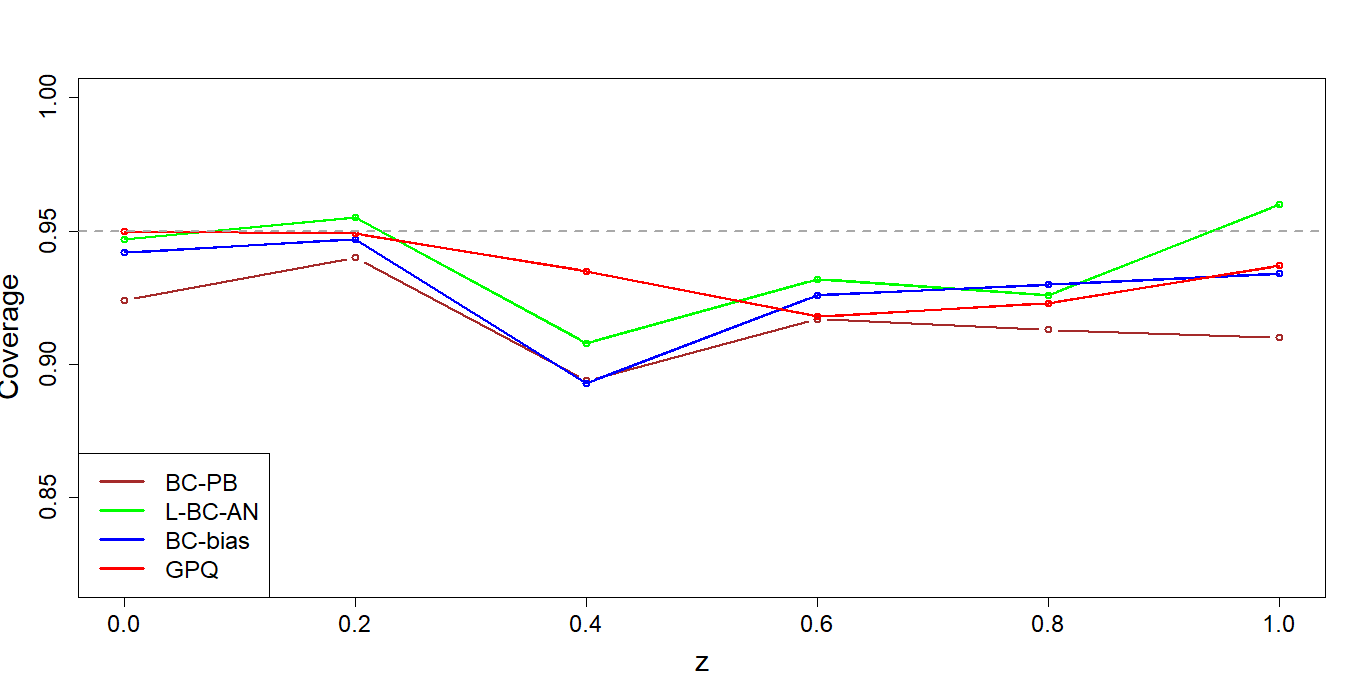}
        \caption*{(50,100)}
    \end{subfigure}
    \begin{subfigure}{0.49\textwidth}
        \includegraphics[width=\linewidth, height=5cm, keepaspectratio]{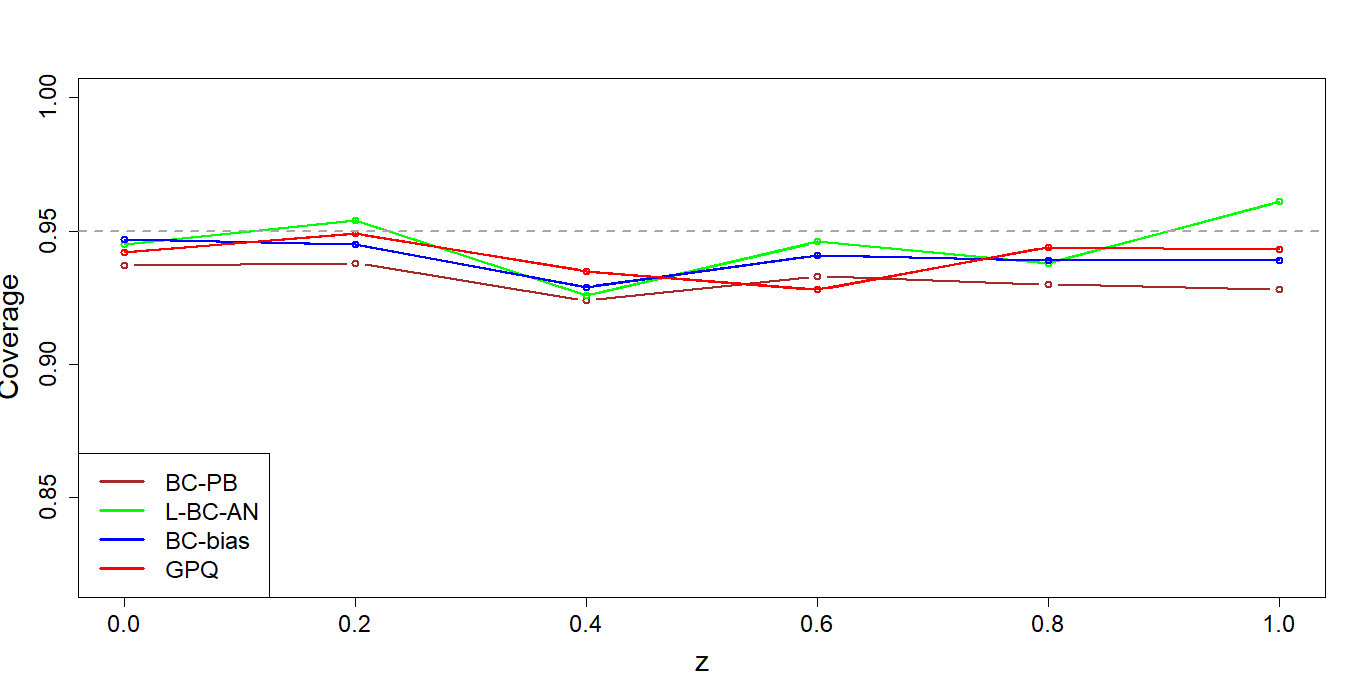}
        \caption*{(100,100)}
    \end{subfigure}
    \begin{subfigure}{0.49\textwidth}
       \includegraphics[width=\linewidth, height=5cm, keepaspectratio]{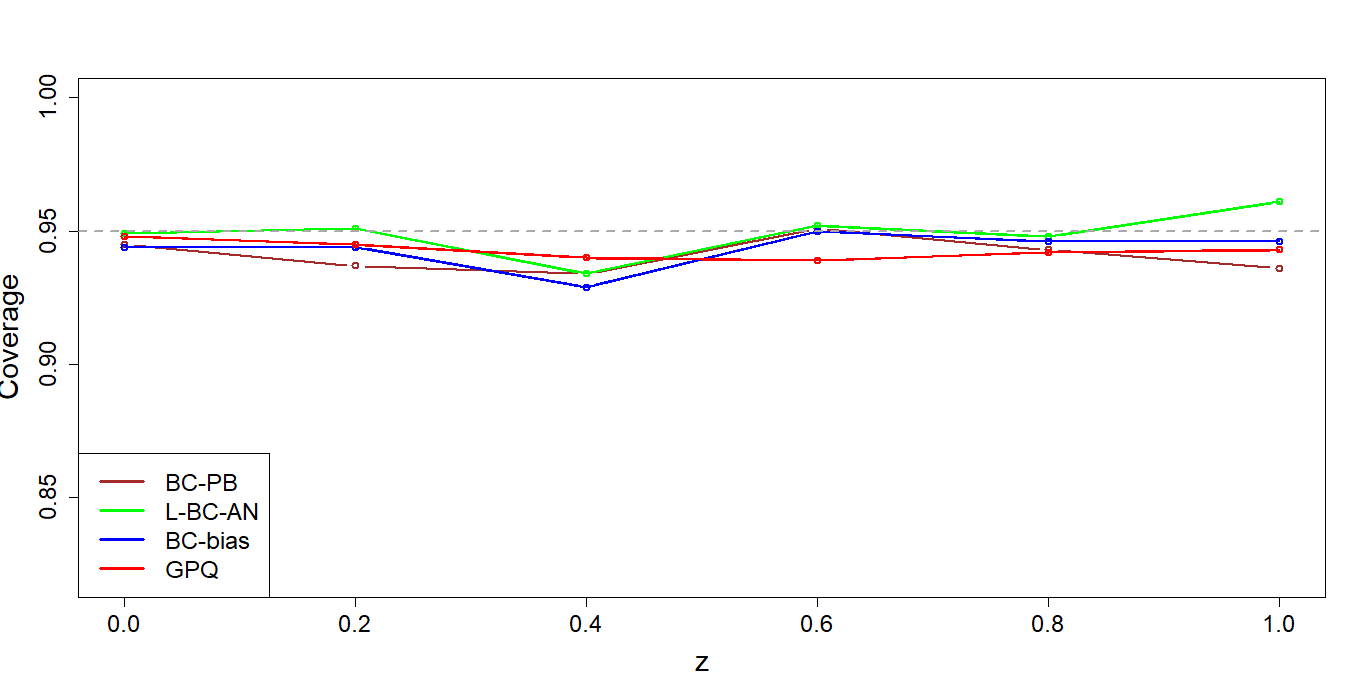}
        \caption*{(200,200)}
    \end{subfigure}
     \caption{Coverages of GPQ and bootstrap methods for the PowerNormal case, scenario 2}
    \label{fig:CPPW2}
\end{figure}

\begin{figure}
    \centering
    \begin{subfigure}{0.49\textwidth}
         \includegraphics[width=\linewidth, height=5cm, keepaspectratio]{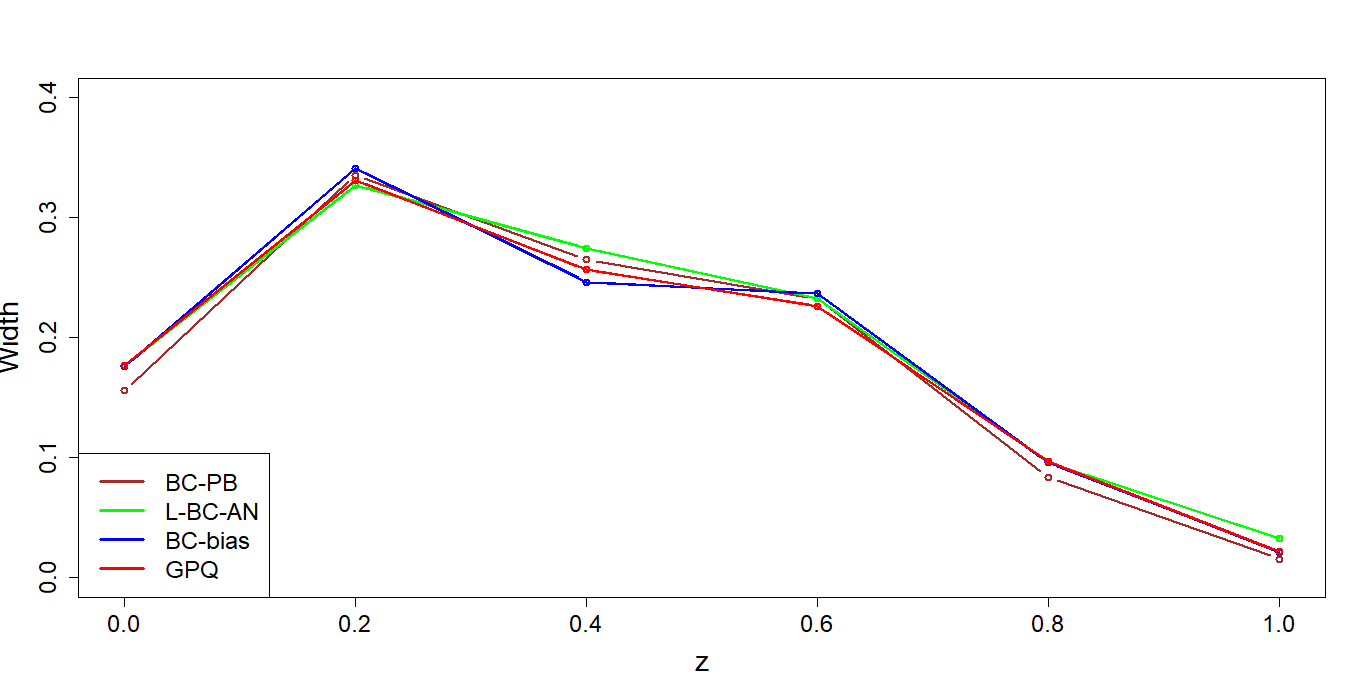}
        \caption*{(50,50)}
    \end{subfigure}
    \begin{subfigure}{0.49\textwidth}
    \includegraphics[width=\linewidth, height=5cm, keepaspectratio]{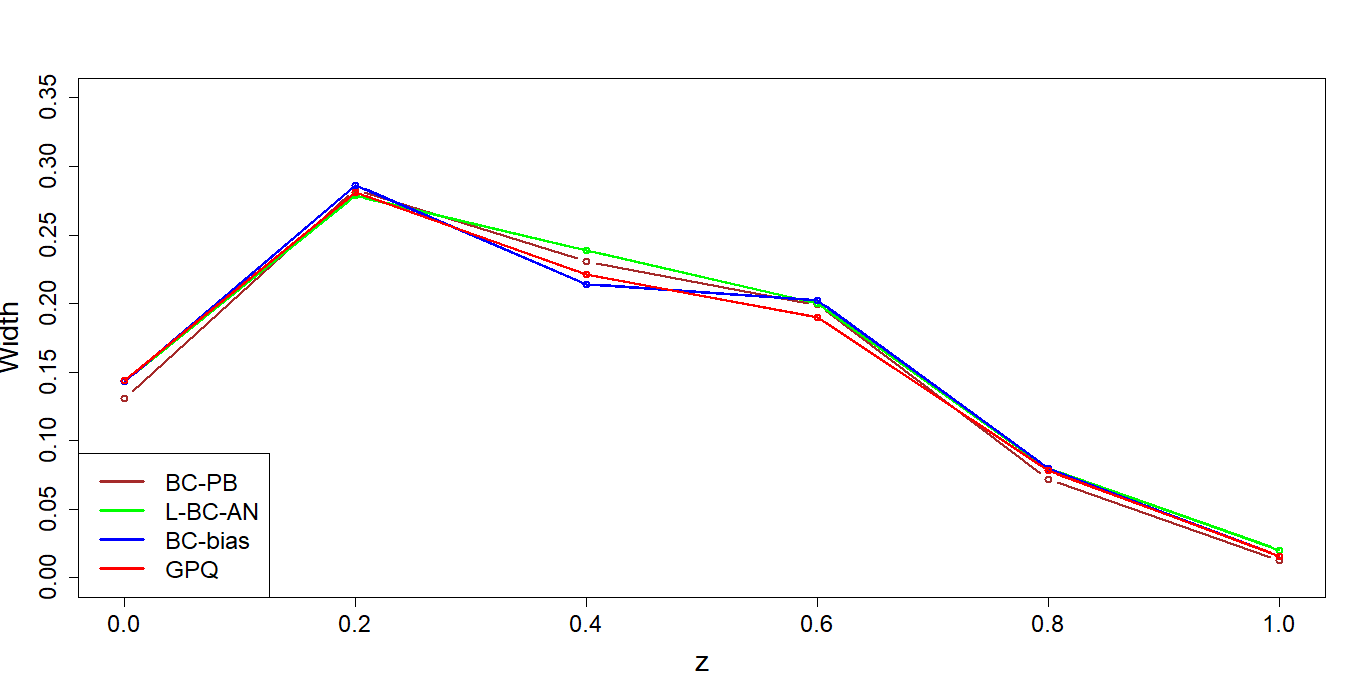}
        \caption*{(50,100)}
    \end{subfigure}
    \begin{subfigure}{0.49\textwidth}
        \includegraphics[width=\linewidth, height=5cm, keepaspectratio]{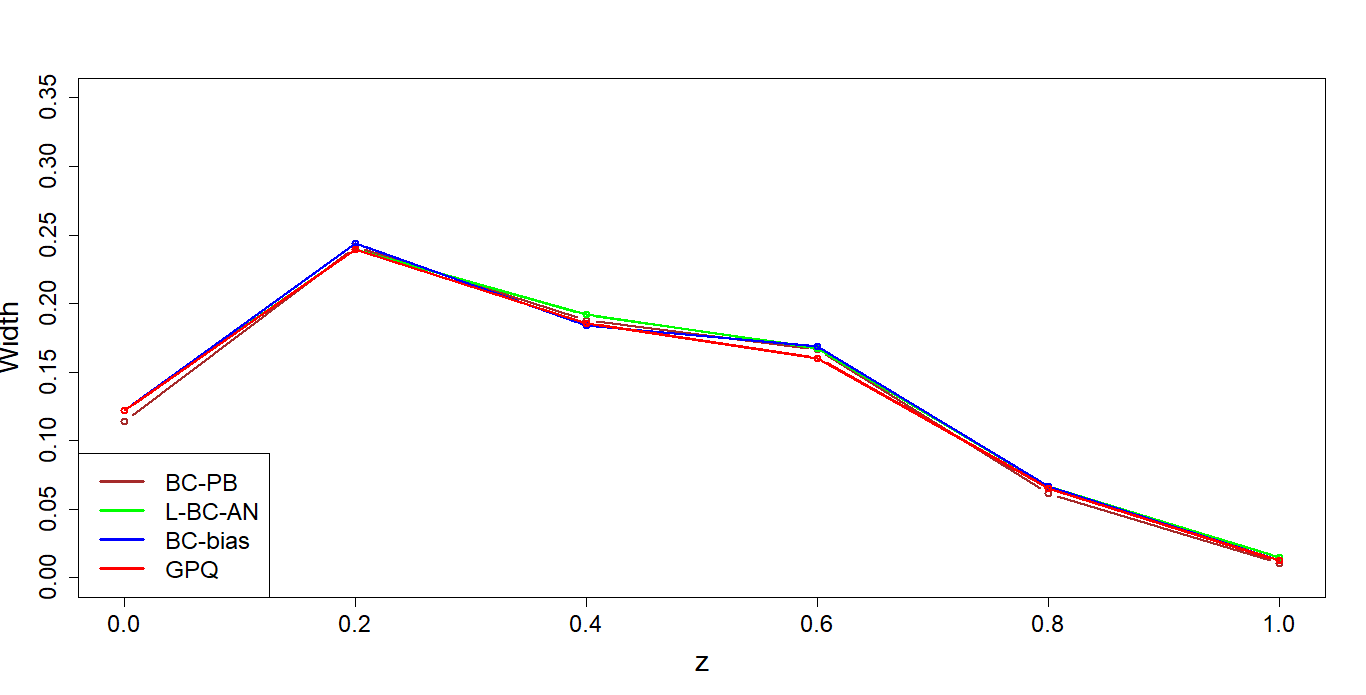}
        \caption*{(100,100)}
    \end{subfigure}
    \begin{subfigure}{0.49\textwidth}
       \includegraphics[width=\linewidth, height=5cm, keepaspectratio]{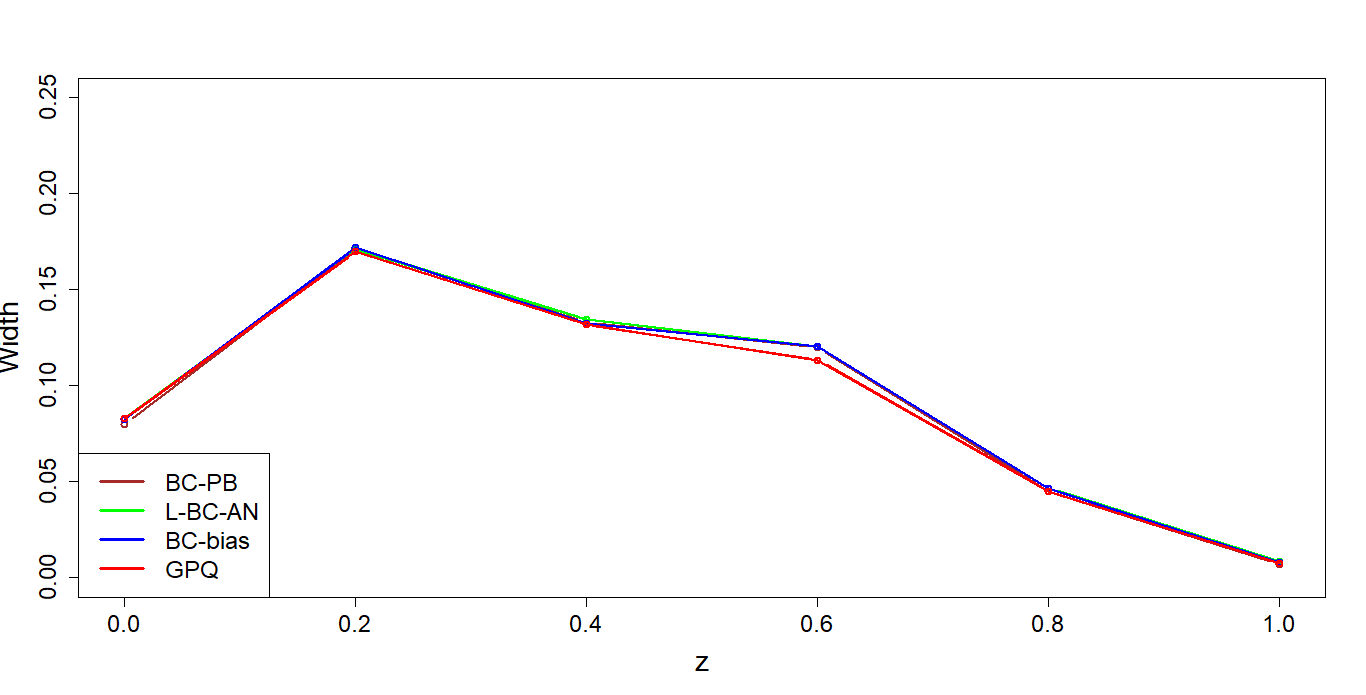}
        \caption*{(200,200)}
    \end{subfigure}
     \caption{Mean interval widths of GPQ and bootstrap methods for the PowerNormal case, scenario 2}
    \label{fig:LENPW2}
\end{figure}
\end{document}